\documentclass[a4paper, amsfonts, superscriptaddress, amsmath, reprint, showkeys, nofootinbib, twoside,floatfix]{revtex4-1}
\usepackage{graphicx}
\usepackage{amsmath}
\usepackage[varg]{txfonts}
\usepackage{xcolor}
\usepackage{epstopdf}
\usepackage{textcomp}
\usepackage{gensymb}
\usepackage{ amssymb }
\usepackage[english]{babel}
\usepackage[utf8]{inputenc}
\usepackage[colorinlistoftodos, color=green!40, prependcaption]{todonotes}
\usepackage[pdftex, pdftitle={Article}, pdfauthor={Author}]{hyperref} % For hyperlinks in the PDF
\renewcommand{\vec}[1]{\mathbf{#1}}% \renewcommand{\vec}[1]{\underline{#1}}
\newcommand{\tens}[1]{\mathbf{#1}}%\newcommand{\tens}[1]{\underline{\underline{#1}}}

\begin{document}
\title{Driven Topological Transitions in Active Nematic Films}

\author{David P. Rivas}
    \affiliation{Department of Physics and Astronomy, Johns Hopkins University, Baltimore, MD, USA, 21218}
\author{Tyler N. Shendruk}
    \affiliation{Interdisciplinary Centre for Mathematical Modelling and Department of Mathematical Sciences, Loughborough University, Loughborough, UK, LE11 3TU}
\author{Robert R. Henry}
    \affiliation{Department of Physics and Astronomy, Johns Hopkins University, Baltimore, MD, USA, 21218}
\author{Daniel H. Reich}
    %\affiliation{Department of Physics and Astronomy, Johns Hopkins University, Baltimore, MD, USA, 21218}
\author{Robert L. Leheny}
    \affiliation{Department of Physics and Astronomy, Johns Hopkins University, Baltimore, MD, USA, 21218}

\date{\today} % Leave empty to omit a date


%\dates{This manuscript was compiled on \today}
%\doi{\url{www.pnas.org/cgi/doi/10.1073/pnas.XXXXXXXXXX}}

\begin{abstract}
The topological properties of many materials are central to their behavior, with the dynamics of topological defects being particularly important to intrinsically out-of-equilibrium, active materials. In this paper, local manipulation of the ordering, dynamics, and topological properties of  microtubule-based extensile active nematic films is demonstrated in a joint experimental and simulation study. Hydrodynamic stresses created by magnetically actuated rotation of disk-shaped colloids in proximity to the films compete with internal stresses in the active nematic, enabling local control of the motion of the +1/2 charge topological defects that are intrinsic to spontaneously turbulent active films. Sufficiently large applied stresses drive the formation of +1 charge topological vortices in the director field through the merger of two +1/2 defects.  The directed motion of the defects is accompanied by ordering of the vorticity and velocity of the active flows within the film that is qualitatively unlike the response of passive viscous films. Many features of the film's response to the disk are captured by Lattice Boltzmann simulations, leading to insight into the anomalous viscoelastic nature of the active nematic. The topological vortex formation is accompanied by a rheological instability in the film that leads to significant increase in the flow velocities.  Comparison of the velocity profile in vicinity of the vortex with fluid-dynamics calculations provides an estimate of film viscosity.
\end{abstract}

\maketitle

Topological defects in the ordered states of physical systems can play critical roles in determining their properties.  In condensed matter, examples include the magnetic flux lines that penetrate type-II superconductors, dislocations in crystalline solids, and singular regions in the magnetization of ferromagnetic thin films \cite{Abrikosov1988,Chaikin2000,Fert2017}.  These structures, which possess a conserved topological charge, have non-local effects on the order within the systems, whose dynamics can often be described in terms of the defects' particle-like motion and interactions. Further, the ability to control defect behavior is essential for many applications.  

Particularly intriguing topological defect dynamics occur in active nematics, which are fluids composed of rod-like constituents that possess the orientational order of nematic liquid crystals but flow spontaneously due to an internal energy source \cite{Marchetti2013,needleman2017,Saintillan2018}. Examples include cell cultures~\cite{Kawaguchi2017,Saw2017,DellArciprete2018,You2018,yaman2019}, bacteria suspensions~\cite{Dombrowski2004,koch2011,Wioland2013,Elgeti2015,NishiguchiPRE2017},  granular media~\cite{Narayan2007}, and engineered systems composed of dense solutions of microtubule or actin biopolymers adsorbed at oil-water interfaces and driven by molecular motors~\cite{Sanchez2012,Kumar2018}.  
In the microtubule films, the local orientation of the microtubules defines the nematic director field ${\bf n}({\bf r})$. Activity induced by the molecular motors causes the microtubules to slide along one another, creating extensional strains that drive bend instabilities in ${\bf n}({\bf r})$ that lead to the perpetual creation and annihilation of pairs of defects with topological charges $\pm 1/2$. The extensional strains further propel the +1/2 defects, and the dynamics of the resulting ensemble of defects leads to striking turbulence-like flows~\cite{doostmohammadi2017,Tan2019,Lemma2019}. 

In the case of traditional liquid crystalline systems, much of their technological importance derives from the ability to address and reconfigure the order locally and dynamically. Progress toward achieving comparable manipulation of active nematics requires control of the dynamic defects.  While recent experiments have demonstrated controlled, uniform flow of bulk active suspensions of microtubular filaments \cite{Wu2017}, as well as alignment of quasi-two-dimensional (2D) active nematic systems \cite{Guillamat2016B,Guillamat2017}, the ability to couple dynamically to local active nematic behavior has not been realized. Here, we report combined experiments and simulations in which we demonstrate and model local control of the flow properties and director field structure in active nematic films dynamically through their interaction with rotating disk-shaped colloids in proximity to the films.  The hydrodynamic stresses produced by the disks alter the behavior of the active nematic by steering the motion of the topological defects, which allows us to infer anomalous viscoelastic properties of the active films. Most notably, however, we find that above a threshold applied stress, the rotating disk can entrain two +1/2 defects and induce their fusion into a single vortex structure with topological charge +1. In overcoming the innate repulsion between like-charged topological objects, which ordinarily is a critical determining factor in the dynamics of multi-defect states, this defect merger illustrates the degree of control of topological properties that can be achieved in active systems. 

\begin{figure}
  \centering
  \includegraphics[width=.95\linewidth]{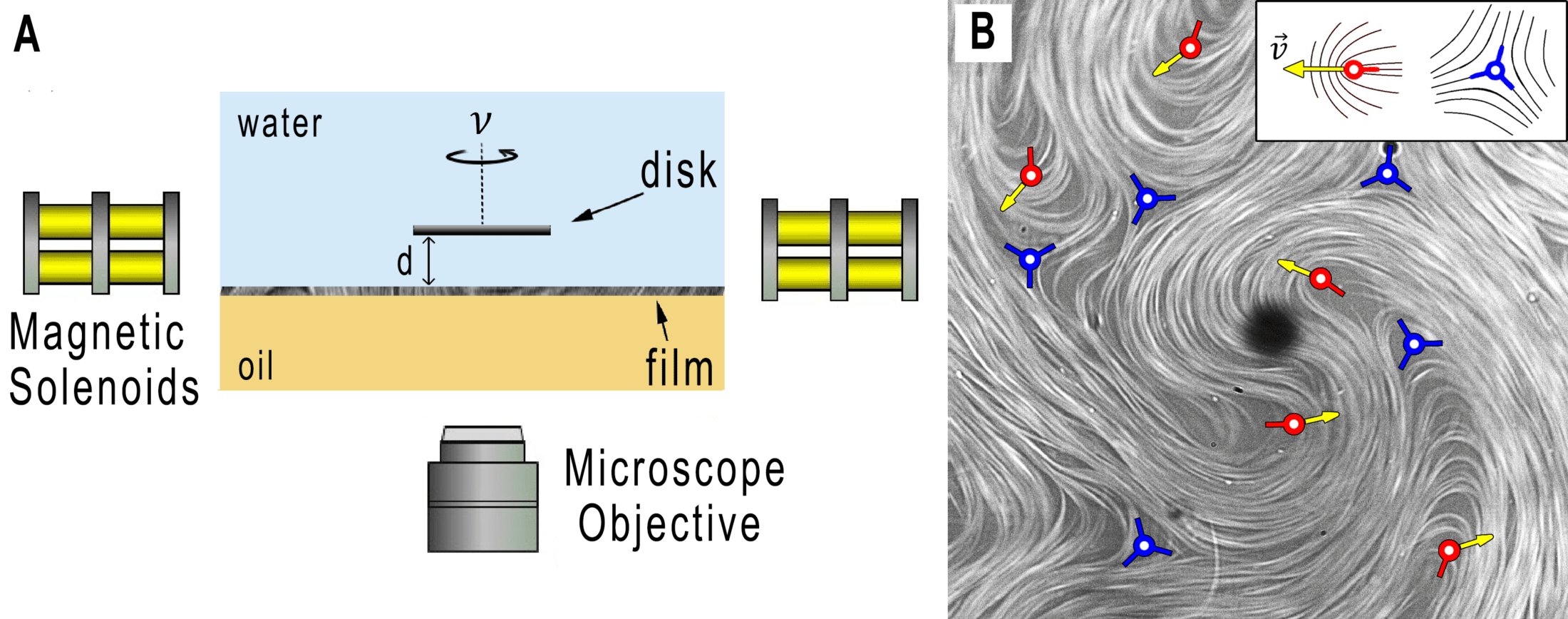}
  \caption {Experimental design({\em A}) A 40-$\mu$m-diameter ferromagnetic nickel disk positioned a height $d = 15-40$ $\mu$m above an active nematic film is rotated at a frequency $\nu$ by a magnetic field created by a set of four pairs of solenoids (two shown) mounted an inverted fluorescence microscope. ({\em B}) Example image of a disk rotating counterclockwise above an active nematic film composed of fluorescently labeled microtubule bundles.  The nematic film is poplulated with $-1/2$ (blue) and $+1/2$ (red) topological defects. The instantaneous velocities of the mobile +1/2 defects are indicated by the yellow arrows, and their orientation vectors are indicated by red arrows. }
  \label{schematic}
\end{figure}

\section{Results and Discussion}

Figure \ref{schematic}{\em A} shows a schematic of the experiments.  Active nematic films composed of a dense layer of microtubule bundles driven by kinesin molecular motors were formed at an oil-water interface \cite{Sanchez2012}, as detailed below in the Materials and Methods section.  As part of the film formation, 40-$\mu$m-diameter ferromagnetic nickel disks, introduced into the aqueous phase, became positioned at a height $d=$ 3 to 8 $\mu$m above the film with the disk faces parallel to the film. The magnetic moments of the disks lie in the disk plane, hence applied  magnetic fields rotating at frequency $\nu$ in the plane of the film caused the disks to rotate about their axes.  When rotated, the disks rose to a greater height above the film.  For instance, at $\nu$ = 80 Hz, the disks typically reached $d=$ 15-40 $\mu$m.  We identify this rise with a possible normal force due to the non-Newtonian character of the aqueous region near the film, which contained a dilute suspension of unadsorbed microtubules.  Figure \ref{schematic}{\em B} shows an image of a rotating disk above an active nematic film where the +1/2 and -1/2 defects are labeled.  The orientation vectors $\hat{\psi}$ of the +1/2 defects~\cite{Vromans2016,tang2017} and the directions of the instantaneous velocities of the defects self-propelled motion, which is approximately antiparallel to $\hat{\psi}$, are further indicated. 

\begin{figure*}
  \centering
  \includegraphics[width=.95\linewidth]{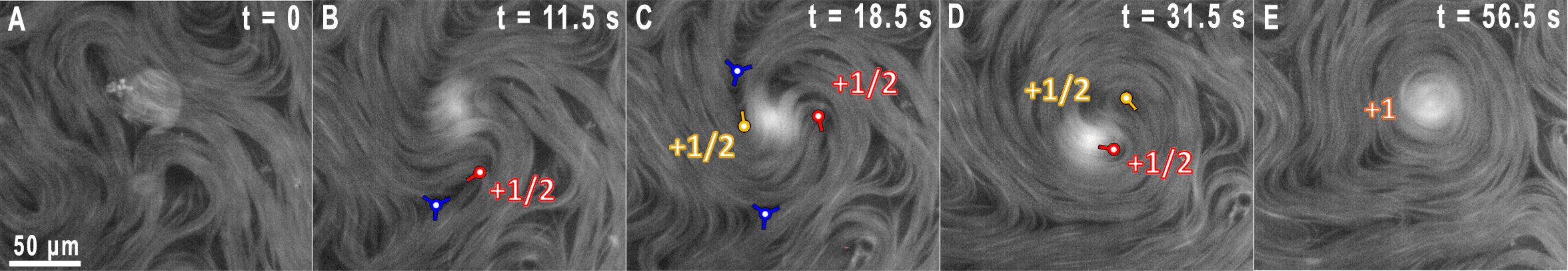}
  \caption {Fluorescence micrographs showing topological vortex formation in an active nematic film due to a disk (appearing as a bright spot) rotating at 80 Hz. ({\em A}) A non-rotating disk sits atop the active nematic film just prior to application of a rotating magnetic field. ({\em B}) A $\pm$1/2 defect pair is created in proximity to the rotating disk. ({\em C}) A second $\pm$1/2 defect pair forms near the disk, while the stresses imposed by the disk cause the positive defects to orient and circle the disk tangentially. ({\em D}) The +1/2 defects spiral in towards the disk center. ({\em E}) The two +1/2 defects fuse to form  a +1 topological vortex, thus conserving the overall charge of the system.}
  \label{vortex_formation}
\end{figure*}

{\bf Topological Vortex Formation and Decay.} When the applied magnetic field was zero, and the disk was not rotating, the disk had no observable effect on the active nematic film; however, the disk was nevertheless carried along with the flow of the film, indicating a strong hydrodynamic coupling between the disks and films.  Consequently, when a disk rotated, it imposed significant hydrodynamic stresses on the film, influencing the nematic order and active flows.  An estimate of these stresses is provided in {\em SI Appendix}, Fig.~S1.  Above a threshold rotation frequency, typically 60-120 Hz, the stresses were sufficient to drive topological transitions in the film, wherein topological vortices with +1 topological charge were created from the fusion of two +1/2 defects. Figure \ref{vortex_formation} shows a series of images of an active nematic film during creation of a topological vortex (see Movie S1).  
%The vortex structure typically had a diameter of 150 to 200 $\mu$m. 
Since total topological charge must be conserved, vortex creation necessitated a change in the population of  1/2-charged defects in the film.  For instance, in Fig.~\ref{vortex_formation}, two +1/2 and -1/2 defect pairs are created in the vicinity of the disk (Fig.~\ref{vortex_formation}{\em B},{\em C}). The +1/2 defects spiral inward toward the disk until they merge to form the +1 vortex defect (Fig.~\ref{vortex_formation}{\em E}), leaving the -1/2 defects behind to eventually annihilate with other +1/2 defects in the film.  The vicinity of the rotating disk was a frequent location for formation of 1/2-charged defect pairs, likely due to the azimuthally directed stresses coupled with the bend instability of the active nematic film, and hence formation of topological vortices commonly involved the merger of +1/2 defects that were created near the disk as in Fig.~\ref{vortex_formation}. However, the +1/2 defects involved in the vortex formation sometimes  originated far from the disk and became captured as they passed in proximity to the disk. 

To explore further the merger of +1/2 defects into a topological object of charge +1, we conducted simulations of active nematic films coupled to rotating disks.  Specifically, we solved the Beris-Edwards formulation of nematohydrodynamics, employing a Lattice-Boltzmann (LB) algorithm for the flow field and finite difference for the orienation tensor~\cite{Marenduzzo2007,foffano2012,Thampi2013,Shendruk2018}. Simulation parameters were chosen to optimize agreement with spatial and temporal correlations in vorticity seen in the experiments, thereby setting the length and time scales for the simulations, as detailed in the SI. The disk in the simulations was subjected to a constant torque to drive rotation and was coupled directly to the active nematic film by a drag coefficient $\zeta_\text{d-f}$ (see  Materials and Methods). 

\begin{figure*}
  \centering
  \includegraphics[width=.95\linewidth]{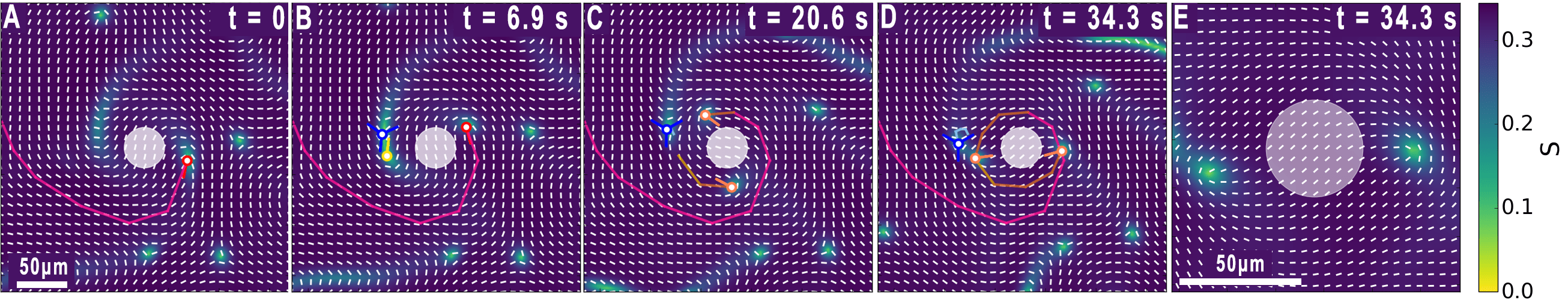}
  \caption {Simulation of the merger of two +1/2 defects into a +1 topological structure by a rotating disk with coupling $\zeta_\text{d-f}=0.03$. ({\em A}) A +1/2 defect (trajectory indicated by red line) begins circling the disk. ({\em B}) A defect pair is created near the disk. ({\em C}) The two +1/2 defects circle the disk. ({\em D}) The +1/2 defects orient in a radial manner. ({\em E}) A magnified view of the +1 topological structure in D.  The white lines display the director field and the color map indicates the scalar order parameter $S$.}
  \label{vortex_formation_sims}
\end{figure*}

Figure~\ref{vortex_formation_sims} illustrates a defect merger event in a simulation (see Movie S2) that mirrors the topological vortex formation in the experiments.  A +1/2 defect was drawn from the surrounding bulk turbulence into the orbit of the rotating disk and proceeded to spiral toward the disk with its polar tail azimuthal (Fig.~\ref{vortex_formation_sims}{\em A}). A second +1/2 defect appeared in the vicinity of the disk by a pair creation event (Fig~\ref{vortex_formation_sims}{\em B}), and the two  +1/2 defects began orbiting the disk on opposites sides from each other.  As they orbited, the defects re-oriented so that each $\hat{\psi}$ pointed radially inward thereby forming a bound pair that constituted a +1 topological complex (Fig~\ref{vortex_formation_sims}{\em C,D}). The +1 topological objects formed in the experiments and simulations hence differed in their core structure.  While in the simulations the core contained two distinguishable +1/2 defects linked by a well defined local director field as seen in Fig.~\ref{vortex_formation_sims}{\em E}, in the experiments the +1/2 defects merged fully to form a point-like core (Fig.~\ref{vortex_formation}{\em E}). We identify this complete merger in the experiments with a nonlinear rheological response of the microtubule-based active nematic films described in more detail below.

\begin{figure*}
  \centering
  \includegraphics[width=.95\linewidth]{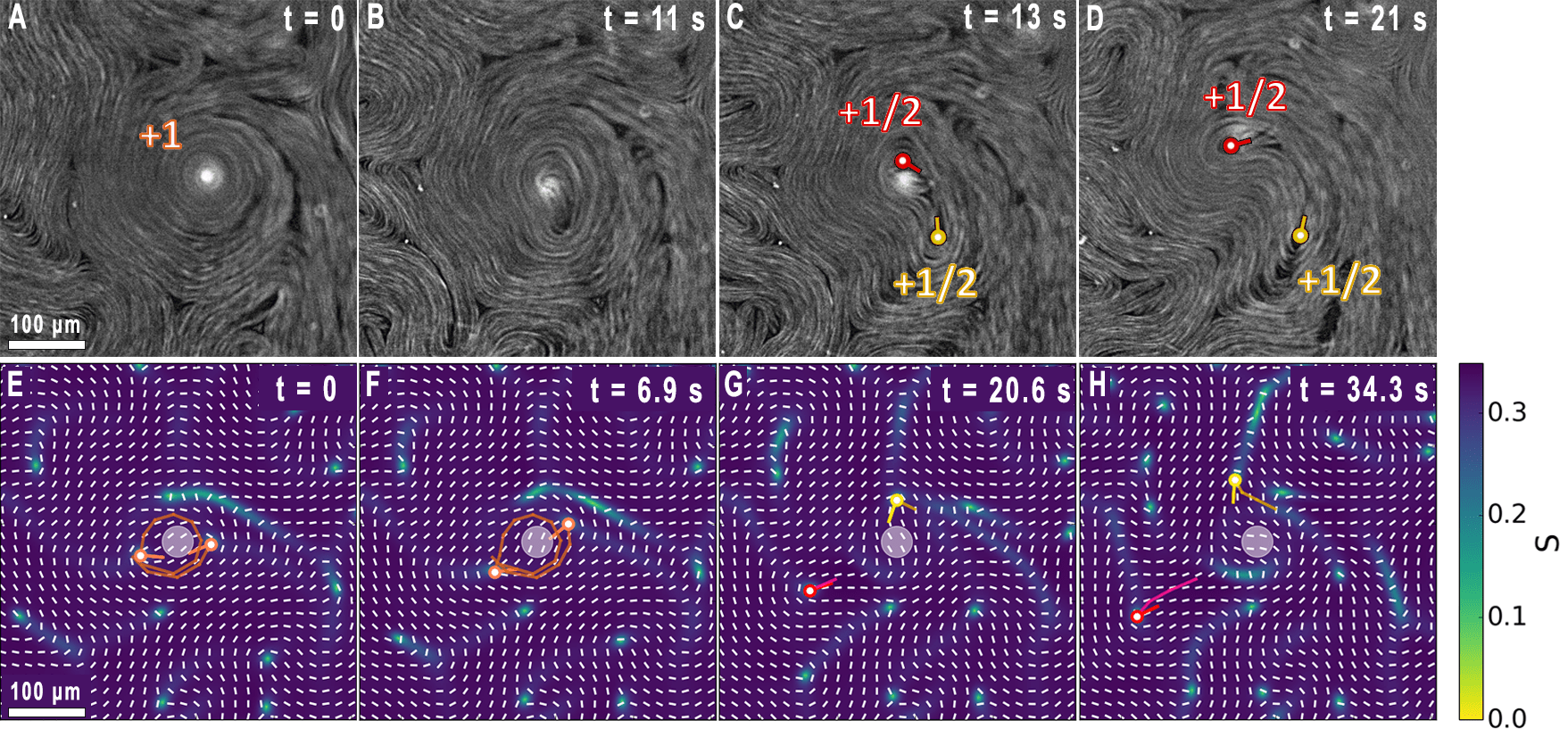}
  \caption {Topological vortex decay. {\em A-D} Fluorescence micrographs showing the +1 topological structure dividing into two +1/2 defects that then propagate away from each other. ({\em A}) shows a +1 topological vortex created by a rotating disk (seen as a bright spot in the image). ({\em B}) shows the initial instability of the azimuthally oriented director about the vortex. A short time later in ({\em C}), two clearly separated +1/2 defects are apparent.  In ({\em D}) the defects continue to propagate away from each other. ({\em E-H}) Simulations (with $\zeta_\text{d-f}=0.03$) showing the decay of a +1 topological structure. The white lines in E-H display the director field and the color map indicates the scalar order parameter $S$.}
  \label{vortex_collapse}
\end{figure*}

Such fusion of like-charge topological defects as in Fig.~\ref{vortex_formation} is unusual in nature.  Like-charge defects typically interact through long-range repulsion~\cite{Vromans2016,tang2017}.  Furthermore, a +1 topological vortex in a nematic has higher energy than two +1/2 defects~\cite{Gennes1993}, so the merger requires external input of energy, in this case provided by the hydrodynamic stress from the disk in conjunction with the activity. 
As high-energy quasiparticles, the topological vortices were hence susceptible to decay. Further, since deformations of the director field in the vicinity of the +1 defect are pure bend, sustaining the configuration required suppressing the activity-driven bend instability intrinsic to the active nematic flms~\cite{SokolovPRX2019}.  In the experiments, the stresses from the rotating disk could stabilize the +1 defect so that it persisted for several minutes; however, fluctuations in the disk position or local nematic order would eventually cause the vortex to decay into a pair of lower-energy $+1/2$ defects that then propagated away from one another (Fig.~\ref{vortex_collapse}{\em A-D} and Movie S3). The +1 topological structures in the simulations decayed similarly (Fig.~\ref{vortex_collapse}{\em E-H}); however, perhaps due to their more complex internal structure, they tended to decay more quickly (see Movie S4). 
%In both the experiments and the simulations, we can see a hydrodynamic instability that is intrinsic to the onset of active turbulence (thampi 2014 and others)~\cite{SokolovPRX2019}. 
 % The rotation of the disk acts against the activity to stabilize this pure bend topology; however, if it is insufficient then the  instability causes the +1 topological complex to decay into two +1/2 defects with a lower free energy cost. 
 
The speed of the defects formed in a vortex decay reached the intrinsic speed of $+1/2$ defects in the active nematic essentially immediately and showed no dependence on their separation, implying the effects of any elastic interactions~\cite{Vromans2016,tang2017} between the defects were overwhelmed by their activity-driven motion~\cite{ShankarPRL2018}  (See {\em SI Appendix}, Fig.~S5).  Once a vortex decayed, the stresses from the rotating disk could with time drive creation of a new vortex, so that the process repeated. 
%{\color{red} Need to quantify with numbers regarding minimum distance and precision of velocity measurement.}

\begin{figure}[h]
  \centering
  \includegraphics[width=.95\linewidth]{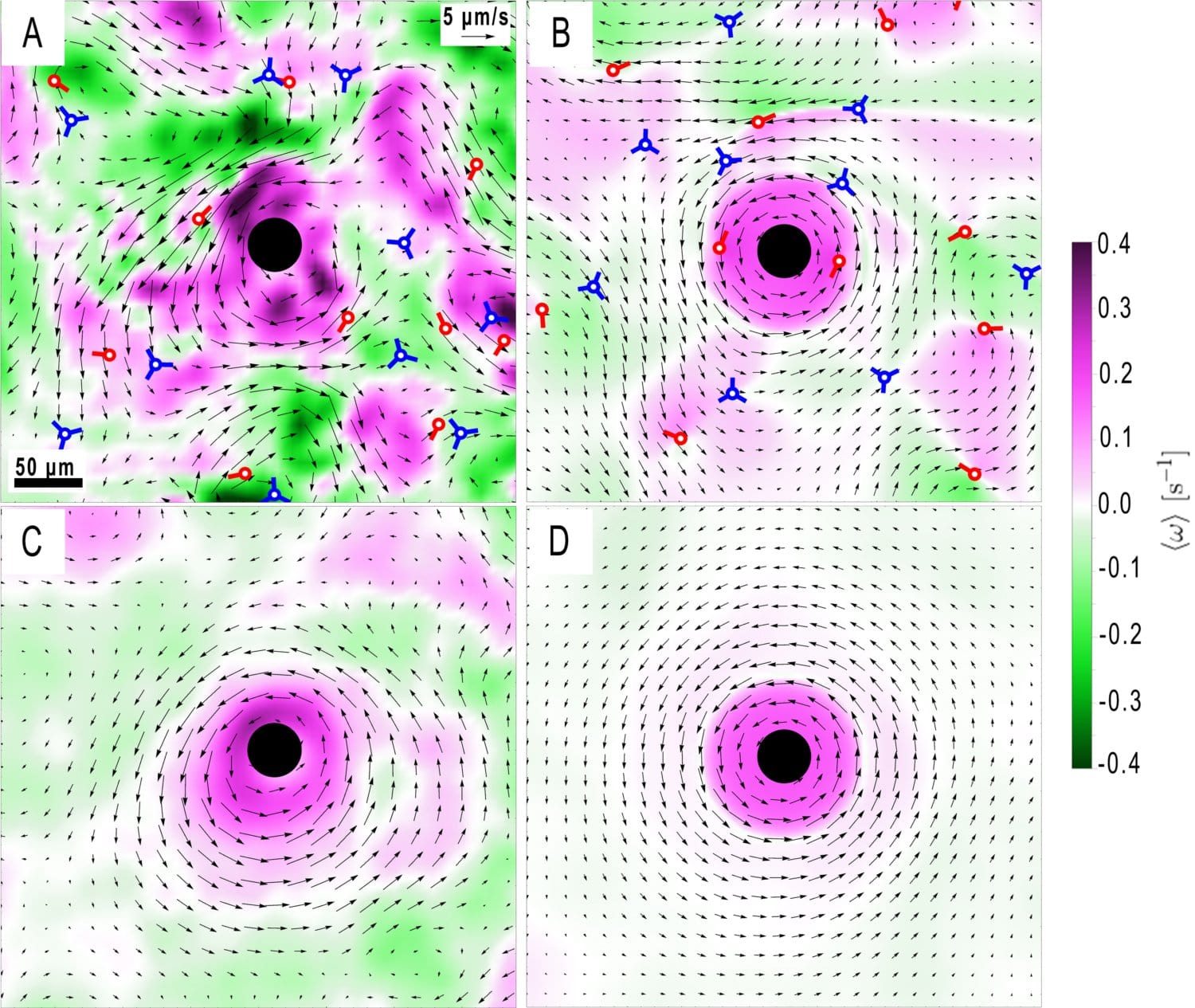}
  \caption {Instantaneous and time-averaged properties of the active nematic flows. ({\em A}) Map of the instantaneous vorticity (color scale) and velocity field (arrows) in an active nematic film in the presence of a disk rotating counterclockwise at 120 Hz. Red and blue markers indicated the +1/2 and -1/2 defects, respectively. ({\em B}) The same instantaneous quantities taken from LB simulation with coupling $\zeta_\text{d-f}=0.07$. ({\em C}) Map of the vorticity and velocity averaged over 300 s in the experiment. ({\em C}) Corresponding time-averaged quantities in the simulation.}
  \label{PIVvelFieldandVort}
\end{figure}

{\bf Effects of Imposed Stress on Active Nematic Structure and Flow.} In addition to the striking formation of a high-energy +1 topological vortex, the hydrodynamics stresses from the rotating disk imposed more subtle effects on the nematic order and active flows.  For example, Fig.~\ref{PIVvelFieldandVort} displays maps generated by analyzing the flows in the vicinity of a disk rotating at 120 Hz during a period in which no +1 topological vortex has formed. (See SI for image analysis details.)  Figure~\ref{PIVvelFieldandVort}{\em A} shows the instantaneous velocity field and the instantaneous vorticity, $\omega =  \vec{\nabla} \times \vec{v}$, in the flow. Active nematics intrinsically possess regions of non-zero flow vorticity~\cite{Lemma2019}, and the regions of positive and negative vorticity in Fig.~\ref{PIVvelFieldandVort}{\em A} reflect this structure. The instantaneous flows in the simulation  show similar structure, as illustrated in Fig.~\ref{PIVvelFieldandVort}{\em B}, and further show how the propagating +1/2 defects are accompanied on either side by regions of clockwise and counter-clockwise local vortices in the flow~\cite{Giomi2015}. In fact, in Fig.~\ref{PIVvelFieldandVort}{\em B} one sees the two +1/2 defects nearest the center of the disk oriented such that their counter-clockwise flow vortices co-rotate with the flow vorticity created by the disk.  
  
Figure~\ref{PIVvelFieldandVort}{\em C,D} displays the same quantities averaged over 300 seconds. Due to the active turbulence of the flow, the structure at a fixed position ordinarily decorrelates on a time scale of approximately 30 seconds (see SI); so in the absence of the rotating disk, both quantities in Fig.~\ref{PIVvelFieldandVort}{\em C,D} would be essentially zero when averaged over 300 seconds.  The non-zero values hence illustrate the ability of the stresses from the rotating disk to compete with the active stress and influence locally the flow of the active nematic.
 
\begin{figure*}
  \centering
  \includegraphics[width=.8\linewidth]{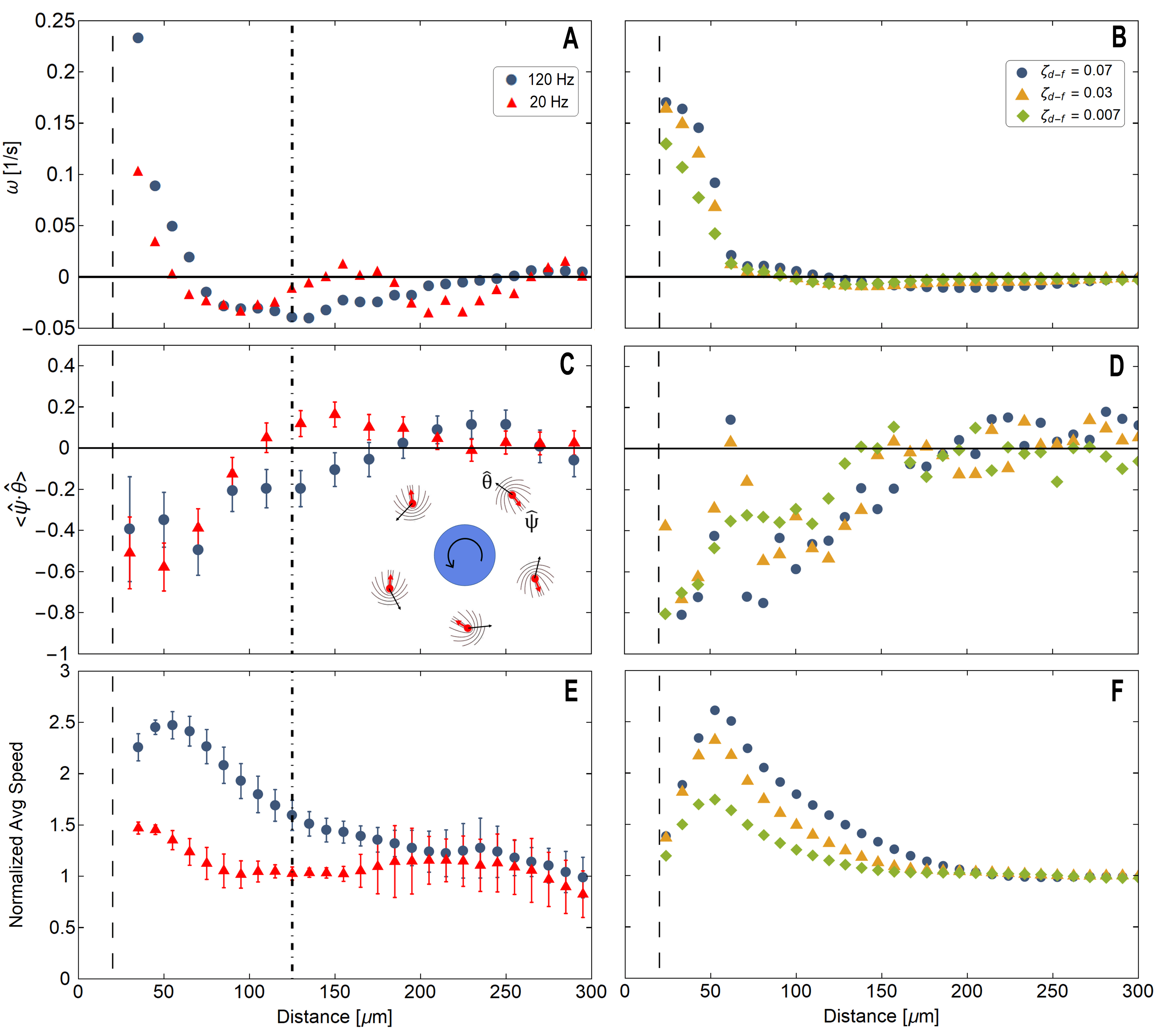}
  \caption {Characterization of the nematic order and flow in the vicinity of a rotating disk in the absence of a +1 topological vortex. {\em A} The time and azimuthally averaged vorticity measured in experiment as a function of distance from the disk center at two disk rotation rates. {\em B} The same quantity obtained in the absence of the +1 topological vortex from simulations at different disk-film couplings. {\em C} The experimental and {\em D} the simulated average azimuthal component of the unit orientation vector, $\hat{\psi}$, of the +1/2 defects as a function of distance from the center of a disk. The negative values near the disk indicate a tendency for the defects to circulate about the disk in its direction of rotation, as shown schematically in the inset in C. {\em E} experimental and {\em F} simulated speed of the film, normalized by the average speed due to the activity far from the disk. The vertical dashed lines indicate the radius of the disk, 20$\mu$m, and the vertical dot-dashed lines indicate the nematic correlation lengths.}
\label{speed}
\end{figure*}

This influence is apparent even at disk rotation rates well below the threshold for creating a topological vortex.  For example, Fig.~\ref{speed} shows a set of quantities characterizing the time-averaged and azimuthally averaged nematic order and flow as a function of distance from the disk center in the absence of a +1 topological object. The experimental data is shown for disk rotation rates of both 120 Hz (during periods when no topological vortex forms) and 20 Hz, a rate below the threshold for vortex formation. Figure~\ref{speed}{\em A} shows the time-averaged flow vorticity which is positive near the disk and negative at intermediate distances. Simulations also show large positive vorticity near the disk and near-zero negative values at larger distance (Fig.~\ref{speed}{\em B}). In the simulations, the region of time-averaged negative vorticity is reproduced only when a small effective friction term~\cite{Thampi2014fric,doostmohammadi2016} is included to account for viscous dissipation due to the thin oil layer below the film (Fig.~\ref{schematic}{\em A}), indicating the importance of such dissipation to the properties of the active flows in experiments.

Orientational ordering of the defects is shown in Fig.~\ref{speed}{\em C}, which displays the time-averaged and azimuthal-averaged component of the +1/2 defect orientation vector $\hat{\psi}$ along the azimuthal direction $\hat{\theta}$ with respect to the disk center. The $+1/2$ defects in close vicinity to the rotating disk tend to orient so that $\left< \hat{\psi} \cdot \hat{\theta}\right> <0$, while farther from the disk $\left< \hat{\psi} \cdot \hat{\theta}\right> >0$.  In the unperturbed active nematic, the extensional flows 
%make the positions of $-1/2$ defects stagnation points, while 
cause the $+1/2$ defects to move anti-parallel to $\hat{\psi}$ as illustrated in the inset of Fig.~\ref{speed}{\em C}.  We thus interpret the region with $\left< \hat{\psi} \cdot \hat{\theta}\right> <0$ as a tendency for the defects to orient so as to facilitate their moving in the net circulating flow seen in Fig.~\ref{PIVvelFieldandVort}{\em C} (see Movie S5).  We further interpret the region farther from the disk in which $\left< \hat{\psi} \cdot \hat{\theta} \right> >0$ as a preference for defects in that region to orient anti-parallel to those closer to the disk since such anti-parallel orientation both minimizes the elastic energy of defect interactions~\cite{Giomi2016,tang2017} and allows favorable constructive overlap of the vorticities in the flows around the defects. Indeed, the tendency for near-neighbor +1/2 defects in active nematics to align anti-parallel has been observed in simulations ~\cite{DeCamp2015} and experiment~\cite{Kumar2018} (see also Fig.~S2). Our simulations in Fig.~\ref{speed}{\em D} show an additional intermediate-distance peak at large effective rotation rates, which represents the slowing of +1/2 defects that are oriented such that they move against the disk's counter-clockwise rotation. This peak due to increased sampling of counter-moving defects is not discernible in the experimental data of Fig.~\ref{speed}{\em C}. 

Figure~\ref{speed}{\em E} displays the time-averaged and azimuthal-average speed of the microtubule bundles in the film, which becomes enhanced over the average activity-induced speeds near the disk. The maximum peak beyond the disk radius is reproduced in LB simulations (Fig.~\ref{speed}{\em F})  by use of an effective hydrodynamic disk size (see Materials and Methods) to account for the effect of the near-disk flows on the film. 

Each experimentally measured quantity in Fig.~\ref{speed} shows similar trends at $\nu =$ 20 Hz and 120 Hz; however, in each case the influence of the disk is larger and extends farther at the higher rate, implying this difference in influence is a factor in whether or not the stresses from the rotating disk can drive topological vortex formation.  (To set the scale of the range of influence, the nematic correlation length, defined in {\em SI Appendix}, Fig~S3, is demarcated by the dash-dotted line in Fig.~\ref{speed}.) 

\begin{figure}[h]
  \centering
  \includegraphics[width=.95\linewidth]{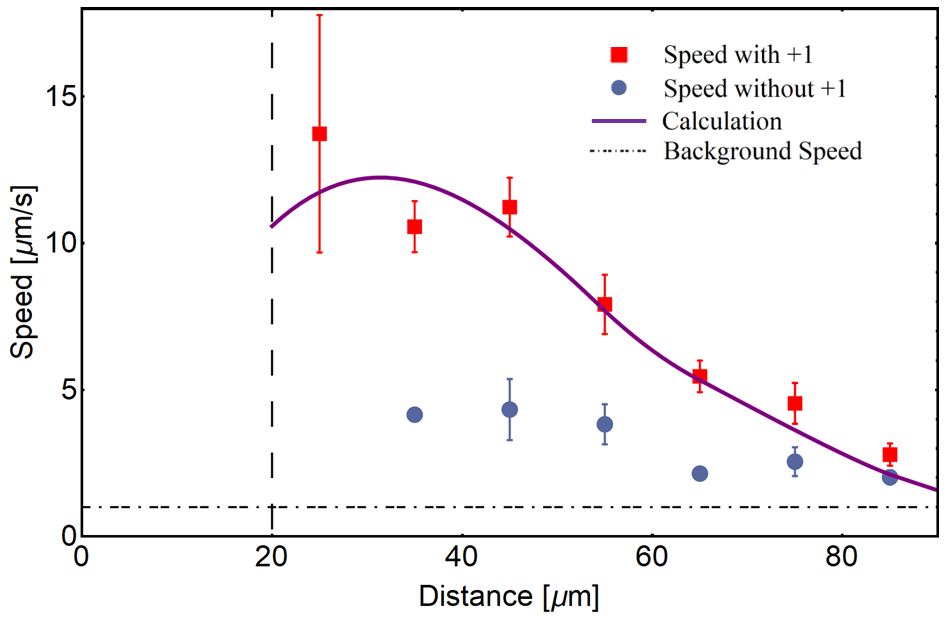}
  \caption {Effect of topological vortex formation on the flow speed of the film as a function of the distance from the center of the rotating disk. The red squares display the speed in the presence of the vortex, while the blue circles display the speed prior to vortex formation. The dashed-dotted line indicates the intrinsic film speed due to the activity. The dashed line shows the disk radius. The solid curve shows the result of a best fit to the measured speed in the presence of the vortex, from which a film viscosity of 1.3 Pa$\cdot$s$\cdot\mu$m was obtained. The disk rotation rate was 80 Hz.}
  \label{vortex_velocity}
\end{figure}

{\bf Shear Thinning of the Active Nematic.} Accompanying the topological vortex formation,  as in Fig.~\ref{vortex_formation}, the velocity of the microtubule bundles in the vicinity of the disk increased significantly, as illustrated in Fig.~\ref{vortex_velocity}, which shows the velocity as a function of distance from the disk center in the presence of a vortex and prior to the vortex formation. The average intrinsic speed of the microtubules far from the disk due to the activity is shown by the dashed line in Fig.~\ref{vortex_velocity}.
Some enhanced flow might be expected in vicinity to the vortex since the nematic director aligns with the shear flow, which is the geometry in which a nematic typically offers the lowest viscous resistance to flow~\cite{Gennes1993}.  However, the large size of the increase in flow velocity that accompanies vortex formation  suggests additional factors.  Specifically, we interpret it as a nonlinear rheological response of the film, in which a local region with reduced viscosity is created in response to the shear stress.  This interpretation is supported by the simulations, where the viscosity terms are not stress dependent and not such comparably enhanced flow velocity is observed.  To analyze this velocity profile, we modeled the experimental conditions in fluid dynamics calculations in COMSOL.  The modeling characterized the low-Reynolds-number hydrodynamics due to a disk positioned at a height $d = 35$ $\mu$m above the film and rotating at $\nu$ = 80 Hz to match the experimental conditions of Fig.~\ref{vortex_velocity}. The model films behaved as quasi-2D fluids in which the velocity profile depended on the 2D viscosity $\eta_{2D}$. By varying $\eta_{2D}$ to optimize agreement between the calculated and measured velocity profiles, we arrived at the solid curve in Fig.~\ref{vortex_velocity}, which corresponds to $\eta_{2D}=$ 1.3 Pa$\cdot$s$\cdot \mu$m. (For details see the Materials and Methods and SI.)  Similar analysis of the  velocity profiles measured around other topological vortices leads to an average viscosity for the films in the range $\eta_{2D} \approx$ 0.7-5 Pa$\cdot$s$\cdot \mu$m. 

 This range of viscosities is approximately two orders of magnitude below the range found in a previous study of microtubule-based active nematics in which the film viscosity  was estimated from the variation in +1/2 defect speed  with  subphase viscosity~\cite{Guillamat2016}. Since we identify the region of the film in vicinity of the vortex with a shear-thinned state, we expect the viscosity to be reduced.  However, the difference between the measured values could also point to the problematic nature of identifying a viscosity in these out-of-equilibrium active systems.  As seen with suspensions of swimming bacteria~\cite{ChenPRL2007,SokolovPRL2009}, and in contrast with conventional Newtonian fluids, different methods of measuring viscosity need not lead to the same result. We note that an additional complication in the interpretation of viscosity in active nematic films arises from the complex and not-yet fully understood coupling between activity, mechanical properties, and topological state of these systems.  

\section{Conclusion}
In this work we have demonstrated a method to couple to and manipulate defect dynamics in an active nematic film.  This capability has allowed us to characterize rheological properties of the active system,  highlighting its anomalous response to external stress. Most dramatically, this response includes a change in the topological character of the nematic as two like-charge $+1/2$ topological defects can be drawn together and caused to merge into a single point defect with topological charge $+1$.  This observation suggests a strategy for  engineering the topology and dynamics of other classes of active matter through appropriate local forcing.

\section{Materials and Methods}
See {\em SI Appendix} for full experimental methods, simulation, and modeling details.
\subsection{Experimental Methods}
\subsubsection{Active Gel}
The ingredients to fabricate the active nematic films were  provided by the Brandeis University Materials Research Science and Engineering Center Biological Materials Facility. Briefly, solutions were prepared by combining appropriate quantities of microtubules, kinesin clusters, polyethylene glycol as a depletant, anti-bleaching agents, ATP, and an ATP regenerating system. The microtubules were labeled with AlexaFluor 647 to enable video tracking of moving bundles using fluorescence microscopy.
\subsubsection{Magnetic Disks}
 Ferromagnetic nickel disks (radius 20 $\mu$m, thickness 300 nm) were fabricated using photolithography methods described elsewhere~\cite{Rovner2012}.  The disks had an in-plane magnetic moment of approximately $1.6 \times 10^{-11} Am^2$ (at an applied field of $3 \times 10^{-3}$ T), as determined from magnetically induced rotations of the disks at a bare oil-water interface.  

 \subsubsection{Sample Preparation}
 Sample cells were composed of two parallel glass slides, one hydrophobically treated and one hydrophilically treated. Thin slices of tape acted as spacers to form channels between the glass surfaces. Oil was inserted into a channel followed immediately by the active nematic mixture containing a dilute dispersion of disks. The chambers were centrifuged, leading to formation at the oil-water interface of a dense film of microtubules with nematic order. Centrifugation of the sample caused the disks to go to the oil-water interface; however, they did not embed in the film but remained in the aqueous phase at a height $d=$ 3 to 8 $\mu$m above the film and oriented with their faces parallel to the film, as depicted schematically in Fig.~\ref{schematic}{\em A}.
 
 \subsubsection{Video Microscopy}
 Observations of the disks and films were made on an inverted microscope (Nikon TE2000) using a Flare CameraLink camera (IOIndustries).  Four pairs of solenoids  mounted on the microscope~\cite{Lee2009} generated rotating magnetic fields of specified magnitude and frequency in the plane of the film, thereby rotating the disks at rates up to $\nu=$ 120 Hz. Image analysis was conducted using custom Python scripts, particle imaging velocimetry (PIV) in Matlab, and ImageJ.
 
 \subsection{Simulations and Numerical Methods}
 
 \subsubsection{Lattice-Boltzmann simulations}
Simulations modelled the microtubule-based active nematic films by solving the equations of motion for incompressible active nematohydrodynamics. The velocity fields obey the Navier-Stokes equations, in which the stress tensor includes viscous, nemato-elastic and active contributions. The active stress is directly proportional to the 3D orientational order parameter tensor~\cite{Ramaswamy2002}, $\tens{Q}=S\left(2\vec{n}\vec{n} -\tens{I}\right)/2$ where $S$ is the scalar order parameter and $\vec{n}$ is the director field. The orientation parameter evolves according to the Beris-Edwards equation of motion and the equations of motion are solved via a hybrid lattice Boltzmann (LB) and finite difference method~\cite{Marenduzzo2007}, with the discrete space and time steps defining the simulation units (su). Simulation units are converted to physical units as discussed in the {\em SI Appendix}, Section 5. 

A constant external torque was applied to a translationally fixed colloidal disk of unit thickness and density $100\times$ that of the microtubule-based film, causing it to rotate with frequency  $\nu$ through the aqueous solution above the film (Fig.~\ref{schematic}) with a large hydrodynamic drag coefficient $\zeta_\text{aq}=10$ su. To account for the experimental height of the disk above the film $d$, which is comparable to the disk  radius $R$, the hydrodynamic coupling of the disk and the film was modelled as a drag force directly below the effective hydrodynamic radius of size $3R$. The coupling strength was determined by a drag coefficient $\zeta_\text{d-f}=\left\{ 0.007, 0.01, 0.03, 0.07 \right\}$ su in Fig.~\ref{speed}. The effective friction in the film due to the proximity of the lower substrate below the oil layer and the associated viscous dissipation was included in the Navier-Stokes equation as a Brinkman term~\cite{Thampi2014fric,doostmohammadi2016} with an effective friction coefficient $\gamma=3\times10^{-4}$ su, which, along with the viscosity $\eta$ and density $\rho$, can be interpreted as a screening length $\sim\sqrt{\eta/\gamma\rho}=112$ $\mu$m.

  \subsubsection{COMSOL modeling}
  To model the velocity profile in vicinity of a topological vortex, we performed hydrodynamic calculations  numerically using COMSOL's Computational Fluid Dynamics (CFD) module to determine the  low-Reynolds-number flows in the film. The modeled geometry consisted of a solid disk rotating above a high-viscosity incompressible fluid film of thickness $h$, diameter $D$, and viscosity $\eta_{f}$, leading to 2D viscosity $\eta_{2D} = h\eta_{f}$, surrounded by bulk, incompressible fluids with viscosity, $\eta_{b}=$ 1 mPa$\cdot$s, matching those of the oil and water in the experiments (see Fig.~\ref{schematic}). 
\newline

 %\showmatmethods{}

\section{ACKNOWLEDGEMENTS} 

We thank Z. Dogic, B. Lemma, and L. Lemma for helpful discussions. We also thank P. Bose for assistance with image analysis. Funding was provided by the NSF (DMR-1610875). We also gratefully acknowledge support from the NSF MRSEC at Brandeis University (DMR-1420382).
This project has received funding (TNS) from the European Research Council (ERC) under the European Union's Horizon 2020 research and innovation programme (grant agreement No. 851196).

\bibliography{topological_transitions_citations_8_29_19}

%merlin.mbs apsrev4-1.bst 2010-07-25 4.21a (PWD, AO, DPC) hacked
%Control: key (0)
%Control: author (72) initials jnrlst
%Control: editor formatted (1) identically to author
%Control: production of article title (-1) disabled
%Control: page (0) single
%Control: year (1) truncated
%Control: production of eprint (0) enabled
\begin{thebibliography}{45}%
\makeatletter
\providecommand \@ifxundefined [1]{%
 \@ifx{#1\undefined}
}%
\providecommand \@ifnum [1]{%
 \ifnum #1\expandafter \@firstoftwo
 \else \expandafter \@secondoftwo
 \fi
}%
\providecommand \@ifx [1]{%
 \ifx #1\expandafter \@firstoftwo
 \else \expandafter \@secondoftwo
 \fi
}%
\providecommand \natexlab [1]{#1}%
\providecommand \enquote  [1]{``#1''}%
\providecommand \bibnamefont  [1]{#1}%
\providecommand \bibfnamefont [1]{#1}%
\providecommand \citenamefont [1]{#1}%
\providecommand \href@noop [0]{\@secondoftwo}%
\providecommand \href [0]{\begingroup \@sanitize@url \@href}%
\providecommand \@href[1]{\@@startlink{#1}\@@href}%
\providecommand \@@href[1]{\endgroup#1\@@endlink}%
\providecommand \@sanitize@url [0]{\catcode `\\12\catcode `\$12\catcode
  `\&12\catcode `\#12\catcode `\^12\catcode `\_12\catcode `\%12\relax}%
\providecommand \@@startlink[1]{}%
\providecommand \@@endlink[0]{}%
\providecommand \url  [0]{\begingroup\@sanitize@url \@url }%
\providecommand \@url [1]{\endgroup\@href {#1}{\urlprefix }}%
\providecommand \urlprefix  [0]{URL }%
\providecommand \Eprint [0]{\href }%
\providecommand \doibase [0]{http://dx.doi.org/}%
\providecommand \selectlanguage [0]{\@gobble}%
\providecommand \bibinfo  [0]{\@secondoftwo}%
\providecommand \bibfield  [0]{\@secondoftwo}%
\providecommand \translation [1]{[#1]}%
\providecommand \BibitemOpen [0]{}%
\providecommand \bibitemStop [0]{}%
\providecommand \bibitemNoStop [0]{.\EOS\space}%
\providecommand \EOS [0]{\spacefactor3000\relax}%
\providecommand \BibitemShut  [1]{\csname bibitem#1\endcsname}%
\let\auto@bib@innerbib\@empty
%</preamble>
\bibitem [{\citenamefont {Abrikosov}(1988)}]{Abrikosov1988}%
  \BibitemOpen
  \bibfield  {author} {\bibinfo {author} {\bibfnamefont {A.}~\bibnamefont
  {Abrikosov}},\ }\href {https://books.google.com/books?id=mKHvAAAAMAAJ} {\emph
  {\bibinfo {title} {Fundamentals of the Theory of Metals}}},\ Fundamentals of
  the Theory of Metals\ (\bibinfo  {publisher} {North-Holland},\ \bibinfo
  {year} {1988})\BibitemShut {NoStop}%
\bibitem [{\citenamefont {Chaikin}\ and\ \citenamefont
  {Lubensky}(2000)}]{Chaikin2000}%
  \BibitemOpen
  \bibfield  {author} {\bibinfo {author} {\bibfnamefont {P.}~\bibnamefont
  {Chaikin}}\ and\ \bibinfo {author} {\bibfnamefont {T.}~\bibnamefont
  {Lubensky}},\ }\href {https://books.google.com/books?id=P9YjNjzr9OIC} {\emph
  {\bibinfo {title} {Principles of Condensed Matter Physics}}}\ (\bibinfo
  {publisher} {Cambridge University Press},\ \bibinfo {year}
  {2000})\BibitemShut {NoStop}%
\bibitem [{\citenamefont {Fert}\ \emph {et~al.}(2017)\citenamefont {Fert},
  \citenamefont {Reyren},\ and\ \citenamefont {Cros}}]{Fert2017}%
  \BibitemOpen
  \bibfield  {author} {\bibinfo {author} {\bibfnamefont {A.}~\bibnamefont
  {Fert}}, \bibinfo {author} {\bibfnamefont {N.}~\bibnamefont {Reyren}}, \ and\
  \bibinfo {author} {\bibfnamefont {V.}~\bibnamefont {Cros}},\ }\href
  {https://doi.org/10.1038/natrevmats.2017.31} {\bibfield  {journal} {\bibinfo
  {journal} {Nature Reviews Materials}\ }\textbf {\bibinfo {volume} {2}},\
  \bibinfo {pages} {17031} (\bibinfo {year} {2017})}\BibitemShut {NoStop}%
\bibitem [{\citenamefont {Marchetti}\ \emph {et~al.}(2013)\citenamefont
  {Marchetti}, \citenamefont {Joanny}, \citenamefont {Ramaswamy}, \citenamefont
  {Liverpool}, \citenamefont {Prost}, \citenamefont {Rao},\ and\ \citenamefont
  {Simha}}]{Marchetti2013}%
  \BibitemOpen
  \bibfield  {author} {\bibinfo {author} {\bibfnamefont {M.~C.}\ \bibnamefont
  {Marchetti}}, \bibinfo {author} {\bibfnamefont {J.~F.}\ \bibnamefont
  {Joanny}}, \bibinfo {author} {\bibfnamefont {S.}~\bibnamefont {Ramaswamy}},
  \bibinfo {author} {\bibfnamefont {T.~B.}\ \bibnamefont {Liverpool}}, \bibinfo
  {author} {\bibfnamefont {J.}~\bibnamefont {Prost}}, \bibinfo {author}
  {\bibfnamefont {M.}~\bibnamefont {Rao}}, \ and\ \bibinfo {author}
  {\bibfnamefont {R.~A.}\ \bibnamefont {Simha}},\ }\href {\doibase
  10.1103/RevModPhys.85.1143} {\bibfield  {journal} {\bibinfo  {journal} {Rev.
  Mod. Phys.}\ }\textbf {\bibinfo {volume} {85}},\ \bibinfo {pages} {1143}
  (\bibinfo {year} {2013})}\BibitemShut {NoStop}%
\bibitem [{\citenamefont {Needleman}\ and\ \citenamefont
  {Dogic}(2017)}]{needleman2017}%
  \BibitemOpen
  \bibfield  {author} {\bibinfo {author} {\bibfnamefont {D.}~\bibnamefont
  {Needleman}}\ and\ \bibinfo {author} {\bibfnamefont {Z.}~\bibnamefont
  {Dogic}},\ }\href {https://doi.org/10.1038/natrevmats.2017.48} {\bibfield
  {journal} {\bibinfo  {journal} {Nature Reviews Materials}\ }\textbf {\bibinfo
  {volume} {2}},\ \bibinfo {pages} {17048} (\bibinfo {year}
  {2017})}\BibitemShut {NoStop}%
\bibitem [{\citenamefont {Saintillan}(2018)}]{Saintillan2018}%
  \BibitemOpen
  \bibfield  {author} {\bibinfo {author} {\bibfnamefont {D.}~\bibnamefont
  {Saintillan}},\ }\href {\doibase 10.1146/annurev-fluid-010816-060049}
  {\bibfield  {journal} {\bibinfo  {journal} {Annual Review of Fluid
  Mechanics}\ }\textbf {\bibinfo {volume} {50}},\ \bibinfo {pages} {563}
  (\bibinfo {year} {2018})},\ \Eprint
  {http://arxiv.org/abs/https://doi.org/10.1146/annurev-fluid-010816-060049}
  {https://doi.org/10.1146/annurev-fluid-010816-060049} \BibitemShut {NoStop}%
\bibitem [{\citenamefont {Kawaguchi}\ \emph {et~al.}(2017)\citenamefont
  {Kawaguchi}, \citenamefont {Kageyama},\ and\ \citenamefont
  {Sano}}]{Kawaguchi2017}%
  \BibitemOpen
  \bibfield  {author} {\bibinfo {author} {\bibfnamefont {K.}~\bibnamefont
  {Kawaguchi}}, \bibinfo {author} {\bibfnamefont {R.}~\bibnamefont {Kageyama}},
  \ and\ \bibinfo {author} {\bibfnamefont {M.}~\bibnamefont {Sano}},\ }\href
  {https://doi.org/10.1038/nature22321} {\bibfield  {journal} {\bibinfo
  {journal} {Nature}\ }\textbf {\bibinfo {volume} {545}},\ \bibinfo {pages}
  {327} (\bibinfo {year} {2017})}\BibitemShut {NoStop}%
\bibitem [{\citenamefont {Saw}\ \emph {et~al.}(2017)\citenamefont {Saw},
  \citenamefont {Doostmohammadi}, \citenamefont {Nier}, \citenamefont
  {Kocgozlu}, \citenamefont {Thampi}, \citenamefont {Toyama}, \citenamefont
  {Marcq}, \citenamefont {Lim}, \citenamefont {Yeomans},\ and\ \citenamefont
  {Ladoux}}]{Saw2017}%
  \BibitemOpen
  \bibfield  {author} {\bibinfo {author} {\bibfnamefont {T.~B.}\ \bibnamefont
  {Saw}}, \bibinfo {author} {\bibfnamefont {A.}~\bibnamefont {Doostmohammadi}},
  \bibinfo {author} {\bibfnamefont {V.}~\bibnamefont {Nier}}, \bibinfo {author}
  {\bibfnamefont {L.}~\bibnamefont {Kocgozlu}}, \bibinfo {author}
  {\bibfnamefont {S.}~\bibnamefont {Thampi}}, \bibinfo {author} {\bibfnamefont
  {Y.}~\bibnamefont {Toyama}}, \bibinfo {author} {\bibfnamefont
  {P.}~\bibnamefont {Marcq}}, \bibinfo {author} {\bibfnamefont {C.~T.}\
  \bibnamefont {Lim}}, \bibinfo {author} {\bibfnamefont {J.~M.}\ \bibnamefont
  {Yeomans}}, \ and\ \bibinfo {author} {\bibfnamefont {B.}~\bibnamefont
  {Ladoux}},\ }\href {https://doi.org/10.1038/nature21718} {\bibfield
  {journal} {\bibinfo  {journal} {Nature}\ }\textbf {\bibinfo {volume} {544}},\
  \bibinfo {pages} {212} (\bibinfo {year} {2017})}\BibitemShut {NoStop}%
\bibitem [{\citenamefont {Dell’Arciprete}\ \emph {et~al.}(2018)\citenamefont
  {Dell’Arciprete}, \citenamefont {Blow}, \citenamefont {Brown},
  \citenamefont {Farrell}, \citenamefont {Lintuvuori}, \citenamefont {McVey},
  \citenamefont {Marenduzzo},\ and\ \citenamefont {Poon}}]{DellArciprete2018}%
  \BibitemOpen
  \bibfield  {author} {\bibinfo {author} {\bibfnamefont {D.}~\bibnamefont
  {Dell’Arciprete}}, \bibinfo {author} {\bibfnamefont {M.~L.}\ \bibnamefont
  {Blow}}, \bibinfo {author} {\bibfnamefont {A.~T.}\ \bibnamefont {Brown}},
  \bibinfo {author} {\bibfnamefont {F.~D.~C.}\ \bibnamefont {Farrell}},
  \bibinfo {author} {\bibfnamefont {J.~S.}\ \bibnamefont {Lintuvuori}},
  \bibinfo {author} {\bibfnamefont {A.~F.}\ \bibnamefont {McVey}}, \bibinfo
  {author} {\bibfnamefont {D.}~\bibnamefont {Marenduzzo}}, \ and\ \bibinfo
  {author} {\bibfnamefont {W.~C.~K.}\ \bibnamefont {Poon}},\ }\href
  {https://doi.org/10.1038/s41467-018-06370-3} {\bibfield  {journal} {\bibinfo
  {journal} {Nature Communications}\ }\textbf {\bibinfo {volume} {9}},\
  \bibinfo {pages} {4190} (\bibinfo {year} {2018})}\BibitemShut {NoStop}%
\bibitem [{\citenamefont {You}\ \emph {et~al.}(2018)\citenamefont {You},
  \citenamefont {Pearce}, \citenamefont {Sengupta},\ and\ \citenamefont
  {Giomi}}]{You2018}%
  \BibitemOpen
  \bibfield  {author} {\bibinfo {author} {\bibfnamefont {Z.}~\bibnamefont
  {You}}, \bibinfo {author} {\bibfnamefont {D.~J.~G.}\ \bibnamefont {Pearce}},
  \bibinfo {author} {\bibfnamefont {A.}~\bibnamefont {Sengupta}}, \ and\
  \bibinfo {author} {\bibfnamefont {L.}~\bibnamefont {Giomi}},\ }\href
  {\doibase 10.1103/PhysRevX.8.031065} {\bibfield  {journal} {\bibinfo
  {journal} {Phys. Rev. X}\ }\textbf {\bibinfo {volume} {8}},\ \bibinfo {pages}
  {031065} (\bibinfo {year} {2018})}\BibitemShut {NoStop}%
\bibitem [{\citenamefont {Yaman}\ \emph {et~al.}(2019)\citenamefont {Yaman},
  \citenamefont {Demir}, \citenamefont {Vetter},\ and\ \citenamefont
  {Kocabas}}]{yaman2019}%
  \BibitemOpen
  \bibfield  {author} {\bibinfo {author} {\bibfnamefont {Y.~I.}\ \bibnamefont
  {Yaman}}, \bibinfo {author} {\bibfnamefont {E.}~\bibnamefont {Demir}},
  \bibinfo {author} {\bibfnamefont {R.}~\bibnamefont {Vetter}}, \ and\ \bibinfo
  {author} {\bibfnamefont {A.}~\bibnamefont {Kocabas}},\ }\href@noop {}
  {\bibfield  {journal} {\bibinfo  {journal} {Nature communications}\ }\textbf
  {\bibinfo {volume} {10}},\ \bibinfo {pages} {2285} (\bibinfo {year}
  {2019})}\BibitemShut {NoStop}%
\bibitem [{\citenamefont {Dombrowski}\ \emph {et~al.}(2004)\citenamefont
  {Dombrowski}, \citenamefont {Cisneros}, \citenamefont {Chatkaew},
  \citenamefont {Goldstein},\ and\ \citenamefont {Kessler}}]{Dombrowski2004}%
  \BibitemOpen
  \bibfield  {author} {\bibinfo {author} {\bibfnamefont {C.}~\bibnamefont
  {Dombrowski}}, \bibinfo {author} {\bibfnamefont {L.}~\bibnamefont
  {Cisneros}}, \bibinfo {author} {\bibfnamefont {S.}~\bibnamefont {Chatkaew}},
  \bibinfo {author} {\bibfnamefont {R.~E.}\ \bibnamefont {Goldstein}}, \ and\
  \bibinfo {author} {\bibfnamefont {J.~O.}\ \bibnamefont {Kessler}},\ }\href
  {\doibase 10.1103/PhysRevLett.93.098103} {\bibfield  {journal} {\bibinfo
  {journal} {Phys. Rev. Lett.}\ }\textbf {\bibinfo {volume} {93}},\ \bibinfo
  {pages} {098103} (\bibinfo {year} {2004})}\BibitemShut {NoStop}%
\bibitem [{\citenamefont {Koch}\ and\ \citenamefont
  {Subramanian}(2011)}]{koch2011}%
  \BibitemOpen
  \bibfield  {author} {\bibinfo {author} {\bibfnamefont {D.~L.}\ \bibnamefont
  {Koch}}\ and\ \bibinfo {author} {\bibfnamefont {G.}~\bibnamefont
  {Subramanian}},\ }\href {\doibase 10.1146/annurev-fluid-121108-145434}
  {\bibfield  {journal} {\bibinfo  {journal} {Annual Review of Fluid
  Mechanics}\ }\textbf {\bibinfo {volume} {43}},\ \bibinfo {pages} {637}
  (\bibinfo {year} {2011})},\ \Eprint
  {http://arxiv.org/abs/https://doi.org/10.1146/annurev-fluid-121108-145434}
  {https://doi.org/10.1146/annurev-fluid-121108-145434} \BibitemShut {NoStop}%
\bibitem [{\citenamefont {Wioland}\ \emph {et~al.}(2013)\citenamefont
  {Wioland}, \citenamefont {Woodhouse}, \citenamefont {Dunkel}, \citenamefont
  {Kessler},\ and\ \citenamefont {Goldstein}}]{Wioland2013}%
  \BibitemOpen
  \bibfield  {author} {\bibinfo {author} {\bibfnamefont {H.}~\bibnamefont
  {Wioland}}, \bibinfo {author} {\bibfnamefont {F.~G.}\ \bibnamefont
  {Woodhouse}}, \bibinfo {author} {\bibfnamefont {J.}~\bibnamefont {Dunkel}},
  \bibinfo {author} {\bibfnamefont {J.~O.}\ \bibnamefont {Kessler}}, \ and\
  \bibinfo {author} {\bibfnamefont {R.~E.}\ \bibnamefont {Goldstein}},\ }\href
  {https://link.aps.org/doi/10.1103/PhysRevLett.110.268102} {\bibfield
  {journal} {\bibinfo  {journal} {PRL}\ }\textbf {\bibinfo {volume} {110}},\
  \bibinfo {pages} {268102} (\bibinfo {year} {2013})}\BibitemShut {NoStop}%
\bibitem [{\citenamefont {Elgeti}\ \emph {et~al.}(2015)\citenamefont {Elgeti},
  \citenamefont {Winkler},\ and\ \citenamefont {Gompper}}]{Elgeti2015}%
  \BibitemOpen
  \bibfield  {author} {\bibinfo {author} {\bibfnamefont {J.}~\bibnamefont
  {Elgeti}}, \bibinfo {author} {\bibfnamefont {R.~G.}\ \bibnamefont {Winkler}},
  \ and\ \bibinfo {author} {\bibfnamefont {G.}~\bibnamefont {Gompper}},\ }\href
  {\doibase 10.1088/0034-4885/78/5/056601} {\bibfield  {journal} {\bibinfo
  {journal} {Reports on Progress in Physics}\ }\textbf {\bibinfo {volume}
  {78}},\ \bibinfo {pages} {056601} (\bibinfo {year} {2015})}\BibitemShut
  {NoStop}%
\bibitem [{\citenamefont {Nishiguchi}\ \emph {et~al.}(2017)\citenamefont
  {Nishiguchi}, \citenamefont {Nagai}, \citenamefont {Chat\'e},\ and\
  \citenamefont {Sano}}]{NishiguchiPRE2017}%
  \BibitemOpen
  \bibfield  {author} {\bibinfo {author} {\bibfnamefont {D.}~\bibnamefont
  {Nishiguchi}}, \bibinfo {author} {\bibfnamefont {K.~H.}\ \bibnamefont
  {Nagai}}, \bibinfo {author} {\bibfnamefont {H.}~\bibnamefont {Chat\'e}}, \
  and\ \bibinfo {author} {\bibfnamefont {M.}~\bibnamefont {Sano}},\ }\href
  {\doibase 10.1103/PhysRevE.95.020601} {\bibfield  {journal} {\bibinfo
  {journal} {Phys. Rev. E}\ }\textbf {\bibinfo {volume} {95}},\ \bibinfo
  {pages} {020601} (\bibinfo {year} {2017})}\BibitemShut {NoStop}%
\bibitem [{\citenamefont {Narayan}\ \emph {et~al.}(2007)\citenamefont
  {Narayan}, \citenamefont {Ramaswamy},\ and\ \citenamefont
  {Menon}}]{Narayan2007}%
  \BibitemOpen
  \bibfield  {author} {\bibinfo {author} {\bibfnamefont {V.}~\bibnamefont
  {Narayan}}, \bibinfo {author} {\bibfnamefont {S.}~\bibnamefont {Ramaswamy}},
  \ and\ \bibinfo {author} {\bibfnamefont {N.}~\bibnamefont {Menon}},\ }\href
  {\doibase 10.1126/science.1140414} {\bibfield  {journal} {\bibinfo  {journal}
  {Science}\ }\textbf {\bibinfo {volume} {317}},\ \bibinfo {pages} {105}
  (\bibinfo {year} {2007})},\ \Eprint
  {http://arxiv.org/abs/https://science.sciencemag.org/content/317/5834/105.full.pdf}
  {https://science.sciencemag.org/content/317/5834/105.full.pdf} \BibitemShut
  {NoStop}%
\bibitem [{\citenamefont {Sanchez}\ \emph {et~al.}(2012)\citenamefont
  {Sanchez}, \citenamefont {Chen}, \citenamefont {DeCamp}, \citenamefont
  {Heymann},\ and\ \citenamefont {Dogic}}]{Sanchez2012}%
  \BibitemOpen
  \bibfield  {author} {\bibinfo {author} {\bibfnamefont {T.}~\bibnamefont
  {Sanchez}}, \bibinfo {author} {\bibfnamefont {D.~T.~N.}\ \bibnamefont
  {Chen}}, \bibinfo {author} {\bibfnamefont {S.~J.}\ \bibnamefont {DeCamp}},
  \bibinfo {author} {\bibfnamefont {M.}~\bibnamefont {Heymann}}, \ and\
  \bibinfo {author} {\bibfnamefont {Z.}~\bibnamefont {Dogic}},\ }\href
  {https://doi.org/10.1038/nature11591} {\bibfield  {journal} {\bibinfo
  {journal} {Nature}\ }\textbf {\bibinfo {volume} {491}},\ \bibinfo {pages}
  {431} (\bibinfo {year} {2012})}\BibitemShut {NoStop}%
\bibitem [{\citenamefont {Kumar}\ \emph {et~al.}(2018)\citenamefont {Kumar},
  \citenamefont {Zhang}, \citenamefont {de~Pablo},\ and\ \citenamefont
  {Gardel}}]{Kumar2018}%
  \BibitemOpen
  \bibfield  {author} {\bibinfo {author} {\bibfnamefont {N.}~\bibnamefont
  {Kumar}}, \bibinfo {author} {\bibfnamefont {R.}~\bibnamefont {Zhang}},
  \bibinfo {author} {\bibfnamefont {J.~J.}\ \bibnamefont {de~Pablo}}, \ and\
  \bibinfo {author} {\bibfnamefont {M.~L.}\ \bibnamefont {Gardel}},\ }\href
  {\doibase 10.1126/sciadv.aat7779} {\bibfield  {journal} {\bibinfo  {journal}
  {Science Advances}\ }\textbf {\bibinfo {volume} {4}} (\bibinfo {year}
  {2018}),\ 10.1126/sciadv.aat7779},\ \Eprint
  {http://arxiv.org/abs/https://advances.sciencemag.org/content/4/10/eaat7779.full.pdf}
  {https://advances.sciencemag.org/content/4/10/eaat7779.full.pdf} \BibitemShut
  {NoStop}%
\bibitem [{\citenamefont {Doostmohammadi}\ \emph {et~al.}(2017)\citenamefont
  {Doostmohammadi}, \citenamefont {Shendruk}, \citenamefont {Thijssen},\ and\
  \citenamefont {Yeomans}}]{doostmohammadi2017}%
  \BibitemOpen
  \bibfield  {author} {\bibinfo {author} {\bibfnamefont {A.}~\bibnamefont
  {Doostmohammadi}}, \bibinfo {author} {\bibfnamefont {T.~N.}\ \bibnamefont
  {Shendruk}}, \bibinfo {author} {\bibfnamefont {K.}~\bibnamefont {Thijssen}},
  \ and\ \bibinfo {author} {\bibfnamefont {J.~M.}\ \bibnamefont {Yeomans}},\
  }\href@noop {} {\bibfield  {journal} {\bibinfo  {journal} {Nature
  Communications}\ }\textbf {\bibinfo {volume} {8}},\ \bibinfo {pages} {15326}
  (\bibinfo {year} {2017})}\BibitemShut {NoStop}%
\bibitem [{\citenamefont {Tan}\ \emph {et~al.}(2019)\citenamefont {Tan},
  \citenamefont {Roberts}, \citenamefont {Smith}, \citenamefont {Olvera},
  \citenamefont {Arteaga}, \citenamefont {Fortini}, \citenamefont {Mitchell},\
  and\ \citenamefont {Hirst}}]{Tan2019}%
  \BibitemOpen
  \bibfield  {author} {\bibinfo {author} {\bibfnamefont {A.~J.}\ \bibnamefont
  {Tan}}, \bibinfo {author} {\bibfnamefont {E.}~\bibnamefont {Roberts}},
  \bibinfo {author} {\bibfnamefont {S.~A.}\ \bibnamefont {Smith}}, \bibinfo
  {author} {\bibfnamefont {U.~A.}\ \bibnamefont {Olvera}}, \bibinfo {author}
  {\bibfnamefont {J.}~\bibnamefont {Arteaga}}, \bibinfo {author} {\bibfnamefont
  {S.}~\bibnamefont {Fortini}}, \bibinfo {author} {\bibfnamefont {K.~A.}\
  \bibnamefont {Mitchell}}, \ and\ \bibinfo {author} {\bibfnamefont {L.~S.}\
  \bibnamefont {Hirst}},\ }\href {https://doi.org/10.1038/s41567-019-0600-y}
  {\bibfield  {journal} {\bibinfo  {journal} {Nature Physics}\ }\textbf
  {\bibinfo {volume} {15}},\ \bibinfo {pages} {1033} (\bibinfo {year}
  {2019})}\BibitemShut {NoStop}%
\bibitem [{\citenamefont {Lemma}\ \emph {et~al.}(2019)\citenamefont {Lemma},
  \citenamefont {DeCamp}, \citenamefont {You}, \citenamefont {Giomi},\ and\
  \citenamefont {Dogic}}]{Lemma2019}%
  \BibitemOpen
  \bibfield  {author} {\bibinfo {author} {\bibfnamefont {L.~M.}\ \bibnamefont
  {Lemma}}, \bibinfo {author} {\bibfnamefont {S.~J.}\ \bibnamefont {DeCamp}},
  \bibinfo {author} {\bibfnamefont {Z.}~\bibnamefont {You}}, \bibinfo {author}
  {\bibfnamefont {L.}~\bibnamefont {Giomi}}, \ and\ \bibinfo {author}
  {\bibfnamefont {Z.}~\bibnamefont {Dogic}},\ }\href {\doibase
  10.1039/C8SM01877D} {\bibfield  {journal} {\bibinfo  {journal} {Soft Matter}\
  }\textbf {\bibinfo {volume} {15}},\ \bibinfo {pages} {3264} (\bibinfo {year}
  {2019})}\BibitemShut {NoStop}%
\bibitem [{\citenamefont {Wu}\ \emph {et~al.}(2017)\citenamefont {Wu},
  \citenamefont {Hishamunda}, \citenamefont {Chen}, \citenamefont {DeCamp},
  \citenamefont {Chang}, \citenamefont {Fern{\'a}ndez-Nieves}, \citenamefont
  {Fraden},\ and\ \citenamefont {Dogic}}]{Wu2017}%
  \BibitemOpen
  \bibfield  {author} {\bibinfo {author} {\bibfnamefont {K.-T.}\ \bibnamefont
  {Wu}}, \bibinfo {author} {\bibfnamefont {J.~B.}\ \bibnamefont {Hishamunda}},
  \bibinfo {author} {\bibfnamefont {D.~T.~N.}\ \bibnamefont {Chen}}, \bibinfo
  {author} {\bibfnamefont {S.~J.}\ \bibnamefont {DeCamp}}, \bibinfo {author}
  {\bibfnamefont {Y.-W.}\ \bibnamefont {Chang}}, \bibinfo {author}
  {\bibfnamefont {A.}~\bibnamefont {Fern{\'a}ndez-Nieves}}, \bibinfo {author}
  {\bibfnamefont {S.}~\bibnamefont {Fraden}}, \ and\ \bibinfo {author}
  {\bibfnamefont {Z.}~\bibnamefont {Dogic}},\ }\href {\doibase
  10.1126/science.aal1979} {\bibfield  {journal} {\bibinfo  {journal}
  {Science}\ }\textbf {\bibinfo {volume} {355}} (\bibinfo {year} {2017}),\
  10.1126/science.aal1979},\ \Eprint
  {http://arxiv.org/abs/https://science.sciencemag.org/content/355/6331/eaal1979.full.pdf}
  {https://science.sciencemag.org/content/355/6331/eaal1979.full.pdf}
  \BibitemShut {NoStop}%
\bibitem [{\citenamefont {Guillamat}\ \emph
  {et~al.}(2016{\natexlab{a}})\citenamefont {Guillamat}, \citenamefont
  {Ign{\'e}s-Mullol},\ and\ \citenamefont {Sagu{\'e}s}}]{Guillamat2016B}%
  \BibitemOpen
  \bibfield  {author} {\bibinfo {author} {\bibfnamefont {P.}~\bibnamefont
  {Guillamat}}, \bibinfo {author} {\bibfnamefont {J.}~\bibnamefont
  {Ign{\'e}s-Mullol}}, \ and\ \bibinfo {author} {\bibfnamefont
  {F.}~\bibnamefont {Sagu{\'e}s}},\ }\href {\doibase 10.1073/pnas.1600339113}
  {\bibfield  {journal} {\bibinfo  {journal} {Proceedings of the National
  Academy of Sciences}\ }\textbf {\bibinfo {volume} {113}},\ \bibinfo {pages}
  {5498} (\bibinfo {year} {2016}{\natexlab{a}})},\ \Eprint
  {http://arxiv.org/abs/https://www.pnas.org/content/113/20/5498.full.pdf}
  {https://www.pnas.org/content/113/20/5498.full.pdf} \BibitemShut {NoStop}%
\bibitem [{\citenamefont {Guillamat}\ \emph {et~al.}(2017)\citenamefont
  {Guillamat}, \citenamefont {Ignés-Mullol},\ and\ \citenamefont
  {Sagués}}]{Guillamat2017}%
  \BibitemOpen
  \bibfield  {author} {\bibinfo {author} {\bibfnamefont {P.}~\bibnamefont
  {Guillamat}}, \bibinfo {author} {\bibfnamefont {J.}~\bibnamefont
  {Ignés-Mullol}}, \ and\ \bibinfo {author} {\bibfnamefont {F.}~\bibnamefont
  {Sagués}},\ }\href {https://doi.org/10.1038/s41467-017-00617-1} {\bibfield
  {journal} {\bibinfo  {journal} {Nature Communications}\ }\textbf {\bibinfo
  {volume} {8}},\ \bibinfo {pages} {564} (\bibinfo {year} {2017})}\BibitemShut
  {NoStop}%
\bibitem [{\citenamefont {Vromans}\ and\ \citenamefont
  {Giomi}(2016{\natexlab{a}})}]{Vromans2016}%
  \BibitemOpen
  \bibfield  {author} {\bibinfo {author} {\bibfnamefont {A.~J.}\ \bibnamefont
  {Vromans}}\ and\ \bibinfo {author} {\bibfnamefont {L.}~\bibnamefont
  {Giomi}},\ }\href {\doibase 10.1039/C6SM01146B} {\bibfield  {journal}
  {\bibinfo  {journal} {Soft Matter}\ }\textbf {\bibinfo {volume} {12}},\
  \bibinfo {pages} {6490} (\bibinfo {year} {2016}{\natexlab{a}})}\BibitemShut
  {NoStop}%
\bibitem [{\citenamefont {Tang}\ and\ \citenamefont
  {Selinger}(2017)}]{tang2017}%
  \BibitemOpen
  \bibfield  {author} {\bibinfo {author} {\bibfnamefont {X.}~\bibnamefont
  {Tang}}\ and\ \bibinfo {author} {\bibfnamefont {J.~V.}\ \bibnamefont
  {Selinger}},\ }\href {\doibase 10.1039/C7SM01195D} {\bibfield  {journal}
  {\bibinfo  {journal} {Soft Matter}\ }\textbf {\bibinfo {volume} {13}},\
  \bibinfo {pages} {5481} (\bibinfo {year} {2017})}\BibitemShut {NoStop}%
\bibitem [{\citenamefont {Marenduzzo}\ \emph {et~al.}(2007)\citenamefont
  {Marenduzzo}, \citenamefont {Orlandini}, \citenamefont {Cates},\ and\
  \citenamefont {Yeomans}}]{Marenduzzo2007}%
  \BibitemOpen
  \bibfield  {author} {\bibinfo {author} {\bibfnamefont {D.}~\bibnamefont
  {Marenduzzo}}, \bibinfo {author} {\bibfnamefont {E.}~\bibnamefont
  {Orlandini}}, \bibinfo {author} {\bibfnamefont {M.~E.}\ \bibnamefont
  {Cates}}, \ and\ \bibinfo {author} {\bibfnamefont {J.~M.}\ \bibnamefont
  {Yeomans}},\ }\href {\doibase 10.1103/PhysRevE.76.031921} {\bibfield
  {journal} {\bibinfo  {journal} {Phys. Rev. E}\ }\textbf {\bibinfo {volume}
  {76}},\ \bibinfo {pages} {031921} (\bibinfo {year} {2007})}\BibitemShut
  {NoStop}%
\bibitem [{\citenamefont {Foffano}\ \emph {et~al.}(2012)\citenamefont
  {Foffano}, \citenamefont {Lintuvuori}, \citenamefont {Stratford},
  \citenamefont {Cates},\ and\ \citenamefont {Marenduzzo}}]{foffano2012}%
  \BibitemOpen
  \bibfield  {author} {\bibinfo {author} {\bibfnamefont {G.}~\bibnamefont
  {Foffano}}, \bibinfo {author} {\bibfnamefont {J.~S.}\ \bibnamefont
  {Lintuvuori}}, \bibinfo {author} {\bibfnamefont {K.}~\bibnamefont
  {Stratford}}, \bibinfo {author} {\bibfnamefont {M.~E.}\ \bibnamefont
  {Cates}}, \ and\ \bibinfo {author} {\bibfnamefont {D.}~\bibnamefont
  {Marenduzzo}},\ }\href {\doibase 10.1103/PhysRevLett.109.028103} {\bibfield
  {journal} {\bibinfo  {journal} {Phys. Rev. Lett.}\ }\textbf {\bibinfo
  {volume} {109}},\ \bibinfo {pages} {028103} (\bibinfo {year}
  {2012})}\BibitemShut {NoStop}%
\bibitem [{\citenamefont {Thampi}\ \emph {et~al.}(2013)\citenamefont {Thampi},
  \citenamefont {Golestanian},\ and\ \citenamefont {Yeomans}}]{Thampi2013}%
  \BibitemOpen
  \bibfield  {author} {\bibinfo {author} {\bibfnamefont {S.~P.}\ \bibnamefont
  {Thampi}}, \bibinfo {author} {\bibfnamefont {R.}~\bibnamefont {Golestanian}},
  \ and\ \bibinfo {author} {\bibfnamefont {J.~M.}\ \bibnamefont {Yeomans}},\
  }\href {\doibase 10.1103/PhysRevLett.111.118101} {\bibfield  {journal}
  {\bibinfo  {journal} {Phys. Rev. Lett.}\ }\textbf {\bibinfo {volume} {111}},\
  \bibinfo {pages} {118101} (\bibinfo {year} {2013})}\BibitemShut {NoStop}%
\bibitem [{\citenamefont {Shendruk}\ \emph {et~al.}(2018)\citenamefont
  {Shendruk}, \citenamefont {Thijssen}, \citenamefont {Yeomans},\ and\
  \citenamefont {Doostmohammadi}}]{Shendruk2018}%
  \BibitemOpen
  \bibfield  {author} {\bibinfo {author} {\bibfnamefont {T.~N.}\ \bibnamefont
  {Shendruk}}, \bibinfo {author} {\bibfnamefont {K.}~\bibnamefont {Thijssen}},
  \bibinfo {author} {\bibfnamefont {J.~M.}\ \bibnamefont {Yeomans}}, \ and\
  \bibinfo {author} {\bibfnamefont {A.}~\bibnamefont {Doostmohammadi}},\ }\href
  {\doibase 10.1103/PhysRevE.98.010601} {\bibfield  {journal} {\bibinfo
  {journal} {Phys. Rev. E}\ }\textbf {\bibinfo {volume} {98}},\ \bibinfo
  {pages} {010601} (\bibinfo {year} {2018})}\BibitemShut {NoStop}%
\bibitem [{\citenamefont {de~Gennes}\ and\ \citenamefont
  {Prost}(1993)}]{Gennes1993}%
  \BibitemOpen
  \bibfield  {author} {\bibinfo {author} {\bibfnamefont {P.}~\bibnamefont
  {de~Gennes}}\ and\ \bibinfo {author} {\bibfnamefont {J.}~\bibnamefont
  {Prost}},\ }\href {https://books.google.com/books?id=0Nw-dzWz5agC} {\emph
  {\bibinfo {title} {The Physics of Liquid Crystals}}},\ International Series
  of Monogr\ (\bibinfo  {publisher} {Clarendon Press},\ \bibinfo {year}
  {1993})\BibitemShut {NoStop}%
\bibitem [{\citenamefont {Sokolov}\ \emph {et~al.}(2019)\citenamefont
  {Sokolov}, \citenamefont {Mozaffari}, \citenamefont {Zhang}, \citenamefont
  {de~Pablo},\ and\ \citenamefont {Snezhko}}]{SokolovPRX2019}%
  \BibitemOpen
  \bibfield  {author} {\bibinfo {author} {\bibfnamefont {A.}~\bibnamefont
  {Sokolov}}, \bibinfo {author} {\bibfnamefont {A.}~\bibnamefont {Mozaffari}},
  \bibinfo {author} {\bibfnamefont {R.}~\bibnamefont {Zhang}}, \bibinfo
  {author} {\bibfnamefont {J.~J.}\ \bibnamefont {de~Pablo}}, \ and\ \bibinfo
  {author} {\bibfnamefont {A.}~\bibnamefont {Snezhko}},\ }\href {\doibase
  10.1103/PhysRevX.9.031014} {\bibfield  {journal} {\bibinfo  {journal} {Phys.
  Rev. X}\ }\textbf {\bibinfo {volume} {9}},\ \bibinfo {pages} {031014}
  (\bibinfo {year} {2019})}\BibitemShut {NoStop}%
\bibitem [{\citenamefont {Shankar}\ \emph {et~al.}(2018)\citenamefont
  {Shankar}, \citenamefont {Ramaswamy}, \citenamefont {Marchetti},\ and\
  \citenamefont {Bowick}}]{ShankarPRL2018}%
  \BibitemOpen
  \bibfield  {author} {\bibinfo {author} {\bibfnamefont {S.}~\bibnamefont
  {Shankar}}, \bibinfo {author} {\bibfnamefont {S.}~\bibnamefont {Ramaswamy}},
  \bibinfo {author} {\bibfnamefont {M.~C.}\ \bibnamefont {Marchetti}}, \ and\
  \bibinfo {author} {\bibfnamefont {M.~J.}\ \bibnamefont {Bowick}},\ }\href
  {\doibase 10.1103/PhysRevLett.121.108002} {\bibfield  {journal} {\bibinfo
  {journal} {Phys. Rev. Lett.}\ }\textbf {\bibinfo {volume} {121}},\ \bibinfo
  {pages} {108002} (\bibinfo {year} {2018})}\BibitemShut {NoStop}%
\bibitem [{\citenamefont {Giomi}(2015)}]{Giomi2015}%
  \BibitemOpen
  \bibfield  {author} {\bibinfo {author} {\bibfnamefont {L.}~\bibnamefont
  {Giomi}},\ }\href {https://link.aps.org/doi/10.1103/PhysRevX.5.031003}
  {\bibfield  {journal} {\bibinfo  {journal} {PRX}\ }\textbf {\bibinfo {volume}
  {5}},\ \bibinfo {pages} {031003} (\bibinfo {year} {2015})}\BibitemShut
  {NoStop}%
\bibitem [{\citenamefont {Thampi}\ \emph {et~al.}(2014)\citenamefont {Thampi},
  \citenamefont {Golestanian},\ and\ \citenamefont {Yeomans}}]{Thampi2014fric}%
  \BibitemOpen
  \bibfield  {author} {\bibinfo {author} {\bibfnamefont {S.~P.}\ \bibnamefont
  {Thampi}}, \bibinfo {author} {\bibfnamefont {R.}~\bibnamefont {Golestanian}},
  \ and\ \bibinfo {author} {\bibfnamefont {J.~M.}\ \bibnamefont {Yeomans}},\
  }\href {\doibase 10.1103/PhysRevE.90.062307} {\bibfield  {journal} {\bibinfo
  {journal} {Phys. Rev. E}\ }\textbf {\bibinfo {volume} {90}},\ \bibinfo
  {pages} {062307} (\bibinfo {year} {2014})}\BibitemShut {NoStop}%
\bibitem [{\citenamefont {Doostmohammadi}\ \emph {et~al.}(2016)\citenamefont
  {Doostmohammadi}, \citenamefont {Adamer}, \citenamefont {Thampi},\ and\
  \citenamefont {Yeomans}}]{doostmohammadi2016}%
  \BibitemOpen
  \bibfield  {author} {\bibinfo {author} {\bibfnamefont {A.}~\bibnamefont
  {Doostmohammadi}}, \bibinfo {author} {\bibfnamefont {M.~F.}\ \bibnamefont
  {Adamer}}, \bibinfo {author} {\bibfnamefont {S.~P.}\ \bibnamefont {Thampi}},
  \ and\ \bibinfo {author} {\bibfnamefont {J.~M.}\ \bibnamefont {Yeomans}},\
  }\href@noop {} {\bibfield  {journal} {\bibinfo  {journal} {Nature
  communications}\ }\textbf {\bibinfo {volume} {7}},\ \bibinfo {pages} {10557}
  (\bibinfo {year} {2016})}\BibitemShut {NoStop}%
\bibitem [{\citenamefont {Vromans}\ and\ \citenamefont
  {Giomi}(2016{\natexlab{b}})}]{Giomi2016}%
  \BibitemOpen
  \bibfield  {author} {\bibinfo {author} {\bibfnamefont {A.~J.}\ \bibnamefont
  {Vromans}}\ and\ \bibinfo {author} {\bibfnamefont {L.}~\bibnamefont
  {Giomi}},\ }\href {\doibase 10.1039/C6SM01146B} {\bibfield  {journal}
  {\bibinfo  {journal} {Soft Matter}\ }\textbf {\bibinfo {volume} {12}},\
  \bibinfo {pages} {6490} (\bibinfo {year} {2016}{\natexlab{b}})}\BibitemShut
  {NoStop}%
\bibitem [{\citenamefont {DeCamp}\ \emph {et~al.}(2015)\citenamefont {DeCamp},
  \citenamefont {Redner}, \citenamefont {Baskaran}, \citenamefont {Hagan},\
  and\ \citenamefont {Dogic}}]{DeCamp2015}%
  \BibitemOpen
  \bibfield  {author} {\bibinfo {author} {\bibfnamefont {S.~J.}\ \bibnamefont
  {DeCamp}}, \bibinfo {author} {\bibfnamefont {G.~S.}\ \bibnamefont {Redner}},
  \bibinfo {author} {\bibfnamefont {A.}~\bibnamefont {Baskaran}}, \bibinfo
  {author} {\bibfnamefont {M.~F.}\ \bibnamefont {Hagan}}, \ and\ \bibinfo
  {author} {\bibfnamefont {Z.}~\bibnamefont {Dogic}},\ }\href
  {https://doi.org/10.1038/nmat4387} {\bibfield  {journal} {\bibinfo  {journal}
  {Nature Materials}\ }\textbf {\bibinfo {volume} {14}},\ \bibinfo {pages}
  {1110} (\bibinfo {year} {2015})}\BibitemShut {NoStop}%
\bibitem [{\citenamefont {Guillamat}\ \emph
  {et~al.}(2016{\natexlab{b}})\citenamefont {Guillamat}, \citenamefont
  {Ignés-Mullol}, \citenamefont {Shankar}, \citenamefont {Marchetti},\ and\
  \citenamefont {Sagués}}]{Guillamat2016}%
  \BibitemOpen
  \bibfield  {author} {\bibinfo {author} {\bibfnamefont {P.}~\bibnamefont
  {Guillamat}}, \bibinfo {author} {\bibfnamefont {J.}~\bibnamefont
  {Ignés-Mullol}}, \bibinfo {author} {\bibfnamefont {S.}~\bibnamefont
  {Shankar}}, \bibinfo {author} {\bibfnamefont {M.~C.}\ \bibnamefont
  {Marchetti}}, \ and\ \bibinfo {author} {\bibfnamefont {F.}~\bibnamefont
  {Sagués}},\ }\href {https://link.aps.org/doi/10.1103/PhysRevE.94.060602}
  {\bibfield  {journal} {\bibinfo  {journal} {PRE}\ }\textbf {\bibinfo {volume}
  {94}},\ \bibinfo {pages} {060602} (\bibinfo {year}
  {2016}{\natexlab{b}})}\BibitemShut {NoStop}%
\bibitem [{\citenamefont {Chen}\ \emph {et~al.}(2007)\citenamefont {Chen},
  \citenamefont {Lau}, \citenamefont {Hough}, \citenamefont {Islam},
  \citenamefont {Goulian}, \citenamefont {Lubensky},\ and\ \citenamefont
  {Yodh}}]{ChenPRL2007}%
  \BibitemOpen
  \bibfield  {author} {\bibinfo {author} {\bibfnamefont {D.~T.~N.}\
  \bibnamefont {Chen}}, \bibinfo {author} {\bibfnamefont {A.~W.~C.}\
  \bibnamefont {Lau}}, \bibinfo {author} {\bibfnamefont {L.~A.}\ \bibnamefont
  {Hough}}, \bibinfo {author} {\bibfnamefont {M.~F.}\ \bibnamefont {Islam}},
  \bibinfo {author} {\bibfnamefont {M.}~\bibnamefont {Goulian}}, \bibinfo
  {author} {\bibfnamefont {T.~C.}\ \bibnamefont {Lubensky}}, \ and\ \bibinfo
  {author} {\bibfnamefont {A.~G.}\ \bibnamefont {Yodh}},\ }\href {\doibase
  10.1103/PhysRevLett.99.148302} {\bibfield  {journal} {\bibinfo  {journal}
  {Phys. Rev. Lett.}\ }\textbf {\bibinfo {volume} {99}},\ \bibinfo {pages}
  {148302} (\bibinfo {year} {2007})}\BibitemShut {NoStop}%
\bibitem [{\citenamefont {Sokolov}\ and\ \citenamefont
  {Aranson}(2009)}]{SokolovPRL2009}%
  \BibitemOpen
  \bibfield  {author} {\bibinfo {author} {\bibfnamefont {A.}~\bibnamefont
  {Sokolov}}\ and\ \bibinfo {author} {\bibfnamefont {I.~S.}\ \bibnamefont
  {Aranson}},\ }\href {\doibase 10.1103/PhysRevLett.103.148101} {\bibfield
  {journal} {\bibinfo  {journal} {Phys. Rev. Lett.}\ }\textbf {\bibinfo
  {volume} {103}},\ \bibinfo {pages} {148101} (\bibinfo {year}
  {2009})}\BibitemShut {NoStop}%
\bibitem [{\citenamefont {Rovner}\ \emph {et~al.}(2012)\citenamefont {Rovner},
  \citenamefont {Borgnia}, \citenamefont {Reich},\ and\ \citenamefont
  {Leheny}}]{Rovner2012}%
  \BibitemOpen
  \bibfield  {author} {\bibinfo {author} {\bibfnamefont {J.~B.}\ \bibnamefont
  {Rovner}}, \bibinfo {author} {\bibfnamefont {D.~S.}\ \bibnamefont {Borgnia}},
  \bibinfo {author} {\bibfnamefont {D.~H.}\ \bibnamefont {Reich}}, \ and\
  \bibinfo {author} {\bibfnamefont {R.~L.}\ \bibnamefont {Leheny}},\ }\href
  {https://link.aps.org/doi/10.1103/PhysRevE.86.041702} {\bibfield  {journal}
  {\bibinfo  {journal} {PRE}\ }\textbf {\bibinfo {volume} {86}},\ \bibinfo
  {pages} {041702} (\bibinfo {year} {2012})}\BibitemShut {NoStop}%
\bibitem [{\citenamefont {Lee}\ \emph {et~al.}(2009)\citenamefont {Lee},
  \citenamefont {Lapointe}, \citenamefont {Reich}, \citenamefont {Stebe},\ and\
  \citenamefont {Leheny}}]{Lee2009}%
  \BibitemOpen
  \bibfield  {author} {\bibinfo {author} {\bibfnamefont {M.~H.}\ \bibnamefont
  {Lee}}, \bibinfo {author} {\bibfnamefont {C.~P.}\ \bibnamefont {Lapointe}},
  \bibinfo {author} {\bibfnamefont {D.~H.}\ \bibnamefont {Reich}}, \bibinfo
  {author} {\bibfnamefont {K.~J.}\ \bibnamefont {Stebe}}, \ and\ \bibinfo
  {author} {\bibfnamefont {R.~L.}\ \bibnamefont {Leheny}},\ }\href {\doibase
  10.1021/la900408y} {\bibfield  {journal} {\bibinfo  {journal} {Langmuir}\
  }\textbf {\bibinfo {volume} {25}},\ \bibinfo {pages} {7976} (\bibinfo {year}
  {2009})}\BibitemShut {NoStop}%
\bibitem [{\citenamefont {Aditi~Simha}\ and\ \citenamefont
  {Ramaswamy}(2002)}]{Ramaswamy2002}%
  \BibitemOpen
  \bibfield  {author} {\bibinfo {author} {\bibfnamefont {R.}~\bibnamefont
  {Aditi~Simha}}\ and\ \bibinfo {author} {\bibfnamefont {S.}~\bibnamefont
  {Ramaswamy}},\ }\href {\doibase 10.1103/PhysRevLett.89.058101} {\bibfield
  {journal} {\bibinfo  {journal} {Phys. Rev. Lett.}\ }\textbf {\bibinfo
  {volume} {89}},\ \bibinfo {pages} {058101} (\bibinfo {year}
  {2002})}\BibitemShut {NoStop}%
\end{thebibliography}%


\begin{thebibliography}{1}

\bibitem{DeCamp2015}
SJ DeCamp, GS Redner, A Baskaran, MF Hagan, Z Dogic, Orientational order of
  motile defects in active nematics.
\newblock {\em\protect\JournalTitle{Nature Materials}} \textbf{14}, 1110
  (2015).

\bibitem{Marenduzzo2007}
D Marenduzzo, E Orlandini, ME Cates, JM Yeomans, Steady-state hydrodynamic
  instabilities of active liquid crystals: Hybrid lattice boltzmann
  simulations.
\newblock {\em\protect\JournalTitle{Phys. Rev. E}} \textbf{76}, 031921 (2007).

\bibitem{thampi2014a}
SP Thampi, R Golestanian, JM Yeomans, Vorticity, defects and correlations in
  active turbulence.
\newblock {\em\protect\JournalTitle{Philosophical Transactions of the Royal
  Society A: Mathematical, Physical and Engineering Sciences}} \textbf{372},
  20130366 (2014).

\bibitem{Thampi2014fric}
SP Thampi, R Golestanian, JM Yeomans, Active nematic materials with substrate
  friction.
\newblock {\em\protect\JournalTitle{Phys. Rev. E}} \textbf{90}, 062307 (2014).

\bibitem{bruus2008}
H Bruus, {\em Theoretical Microfluidics}, Oxford Master Series in Physics.
\newblock (Oxford University Press Oxford), (2008).

\bibitem{denniston2004}
C Denniston, D Marenduzzo, E Orlandini, J Yeomans, Lattice boltzmann algorithm
  for three-dimensional liquid-crystal hydrodynamics.
\newblock {\em\protect\JournalTitle{Philosophical Transactions of the Royal
  Society of London. Series A: Mathematical, Physical and Engineering
  Sciences}} \textbf{362}, 1745--1754 (2004).

\bibitem{Shendruk2018}
TN Shendruk, K Thijssen, JM Yeomans, A Doostmohammadi, Twist-induced crossover
  from two-dimensional to three-dimensional turbulence in active nematics.
\newblock {\em\protect\JournalTitle{Phys. Rev. E}} \textbf{98}, 010601 (2018).

\bibitem{PIVLab}
W Thielicke, EJ Stamhuis, Pivlab - time-resolved digital particle image
  velocimetry tool for matlab (version: 2.00) (2014).

\bibitem{Trackpy}
NKFBRP D.~Allan, T. A.~Caswell, L Uieda, trackpy: Trackpy v0.3.3 (2014).

\end{thebibliography}
\end{document}

%% Comment/remove this line before generating final copy for submission; this will also remove the warning re "Consecutive odd pages found".
%\instructionspage  

%Tyler made a couple of macro's for math so that we can just change notation once if we ever want (rather than combing through the doc)
\renewcommand{\vec}[1]{\mathbf{#1}}% \renewcommand{\vec}[1]{\underline{#1}}
\newcommand{\tens}[1]{\mathbf{#1}}%\newcommand{\tens}[1]{\underline{\underline{#1}}}
\newcommand{\del}{\vec{\nabla}}
\newcommand{\tr}[1]{\mathrm{Tr}\left[#1\right]}

%\maketitle

%% Adds the main heading for the SI text. Comment out this line if you do not have any supporting information text.
%\SItext

\begin{center}
    \fontsize{18pt}{12pt}\selectfont
    \textbf{Supplementary Information}
\end{center}
\vspace{20px}

\section{Sample Preparation}
\subsection{Active Gel Fabrication}
The active gel was fabricated by combining  microtubules, ATP, and a premix solution that contained kinesin clusters, anti-bleaching agents, polyethylene glycol as a depletant, an ATP regeneration system, and buffer, as described in \cite{DeCamp2015}.  The microtubules and premix were provided by Marc Radilla from the Biological Materials Facility at Brandeis University. The final concentration of polyethylene glycol was 0.8$\%$ (w/v) and the final concentration of ATP was typically 1.4 mM.

\subsection{Sample Cell Fabrication}
Sample cells were fabricated from  two glass slides. The  slides  were first cut to approximately 1" $\times$ 0.5" then cleaned in hot water containing 1$\%$ Hellmanex (Hellma) solution and sonicated for $\sim$10 minutes. One slide was made hydrophilic by coating with polyacrylamide, and the other slide was hydrophobically treated using Aquapel (Pittsburgh Glass Works). The hydrophilic treatment involved first sonicating the cleaned glass in ethanol for ten minutes, then rinsing with deionized (DI) water and sonicating in 0.1 M KOH, also for approximately ten minutes. A solution of DI water and 2$\%$ w/v acrylamide was placed under vacuum for at least 15 minutes prior to adding $0.035\%$ tetramethylethylenediamine and 0.7 g/L ammonium persulfate. The slides were soaked in a solution of 98.5$\%$ ethanol, 1$\%$ acetic acid, and 0.5$\%$ 3-(trimethoxysilyl) propylmethacrylate for roughly 15 minutes, rinsed again, and then soaked overnight in the acrylamide solution. The slides were rinsed in DI water and dried prior to use. The hydrophobic treatment involved coating the glass in Aquapel for approximately one minute then rinsing with DI water and drying. Thin strips of double-sided tape were placed on the surface of the hydrophobic slides to create two channels of approximate dimensions 2 cm $\times$ 0.2 cm in which two active nematic films were created. The hydrophilic slides were then laid on top and pressed down onto the tape. 
%The excess ends of the tape were cut off 

Fluorinated oil (HFE 7500, RAN Biotechnologies) was pipetted into a channel via capillary action. The active gel containing a dilute dispersion of magnetic disks was immediately pipetted into the same channel displacing the oil except for a thin layer adjacent to the hydrophobic surface. The sides of the cell were then sealed with Norland  Optical Adhesive 81 (Norland Products), and the cell was centrifuged for 10-15 minutes at 800 rpm (103 g) in order to drive the microtubules and disks to the interface.

\section{Estimate of Stress on Active Nematic Film from Rotating Disk}
\begin{figure}[h]
\centering
\includegraphics[width=0.6\textwidth]{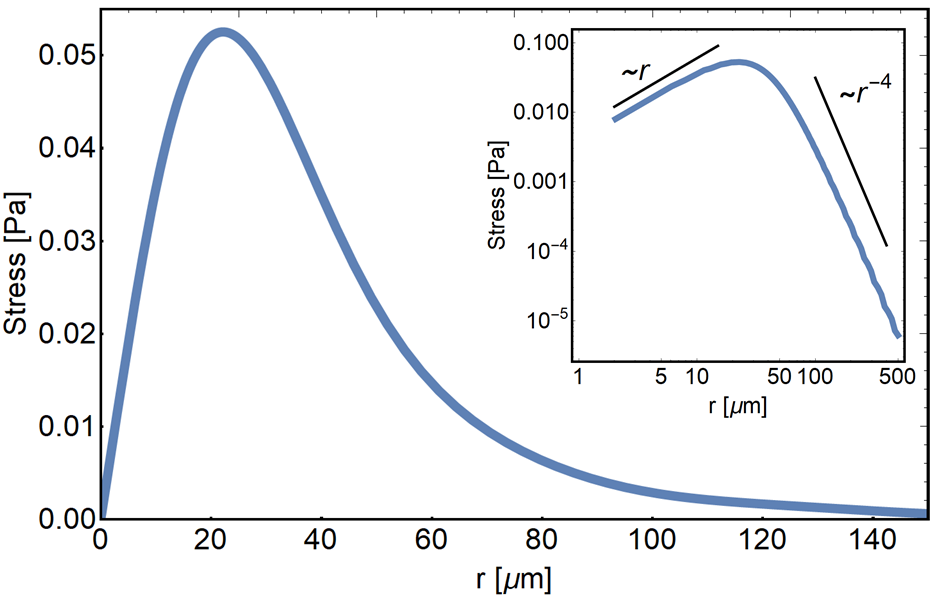}
\caption {Computed stress on a film surface due to shear flow in the adjacent water created by a 40-$\mu$m-diameter disk rotating at 80 Hz at a height of 35 $\mu$m above the film, plotted as a function of distance $r$ from the center of the disk.  The inset shows the stress on a log-log scale.}
\label{stress}
\end{figure}

Estimates of the hydrodynamic stress imposed on the film by a disk rotating in the water layer above the film were performed using COMSOL's Computational Fluid Dynamics module.  As an example, Fig.~\ref{stress} displays the stress at the film surface as a function of horizontal distance $r$ from the disk center  computed for a 40 $\mu$m-diameter disk positioned 35 $\mu$m above the film and rotating at 80 Hz.  The surface of the rotating disk imposed a no-slip boundary condition creating a shear flow in the water, and due to the large difference in viscosities between the water and film, we approximated the film as a stationary surface that also imposed a no-slip boundary condition on the water's flow. As shown in the inset, the stress decays as $r^{-4}$ at large $r$, while at small values of $r$ corresponding to the film directly under the disk, the stress increases linearly with $r$. 

\section{+1/2-Defect Orientation Pair Correlation Function}
\begin{figure}[h]
\centering
\includegraphics[width=0.6\textwidth]{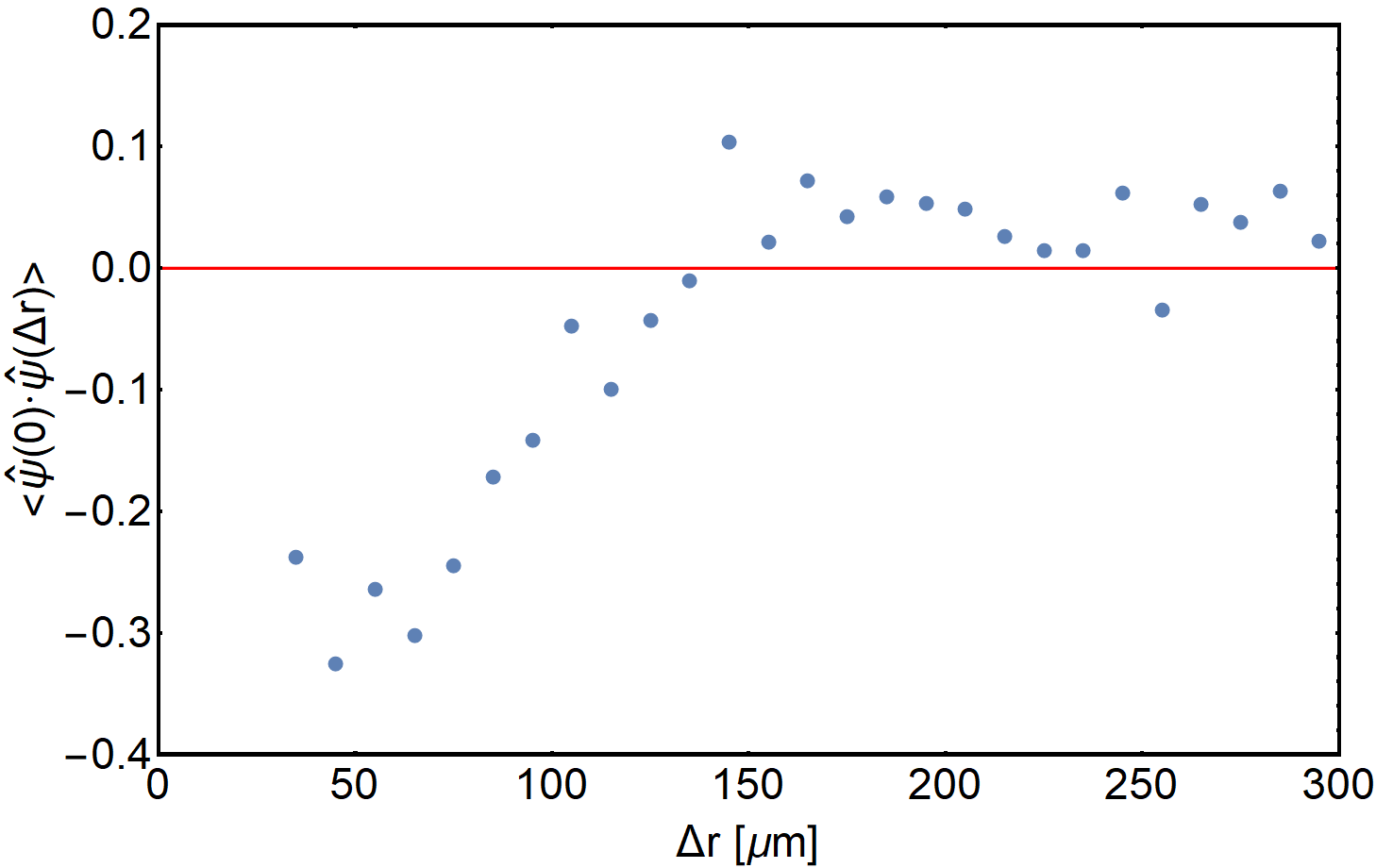}
\caption {The orientation correlation function of +1/2 defects as a function of defect separation $\Delta r$. The negative values at short separations indicate that nearby +1/2 defects tend to be anti-aligned.}
\label{defectcorr}
\end{figure}

Figure~\ref{defectcorr} displays the orientation correlation function between +1/2 defects in an active nematic film as a function of the separation $\Delta r$ of the defects.  The correlation function is defined as the average value of the dot product of the orientation unit vectors, $\langle\hat{\psi}(0)\cdot\hat{\psi}(\Delta r)\rangle$.  Nearby +1/2 defects tend to be anti-aligned, as seen by the negative values of the correlation function at small separations.

\section{Determination of the Nematic Correlation Length}
\begin{figure}[h]
\centering
\includegraphics[width=0.6\textwidth]{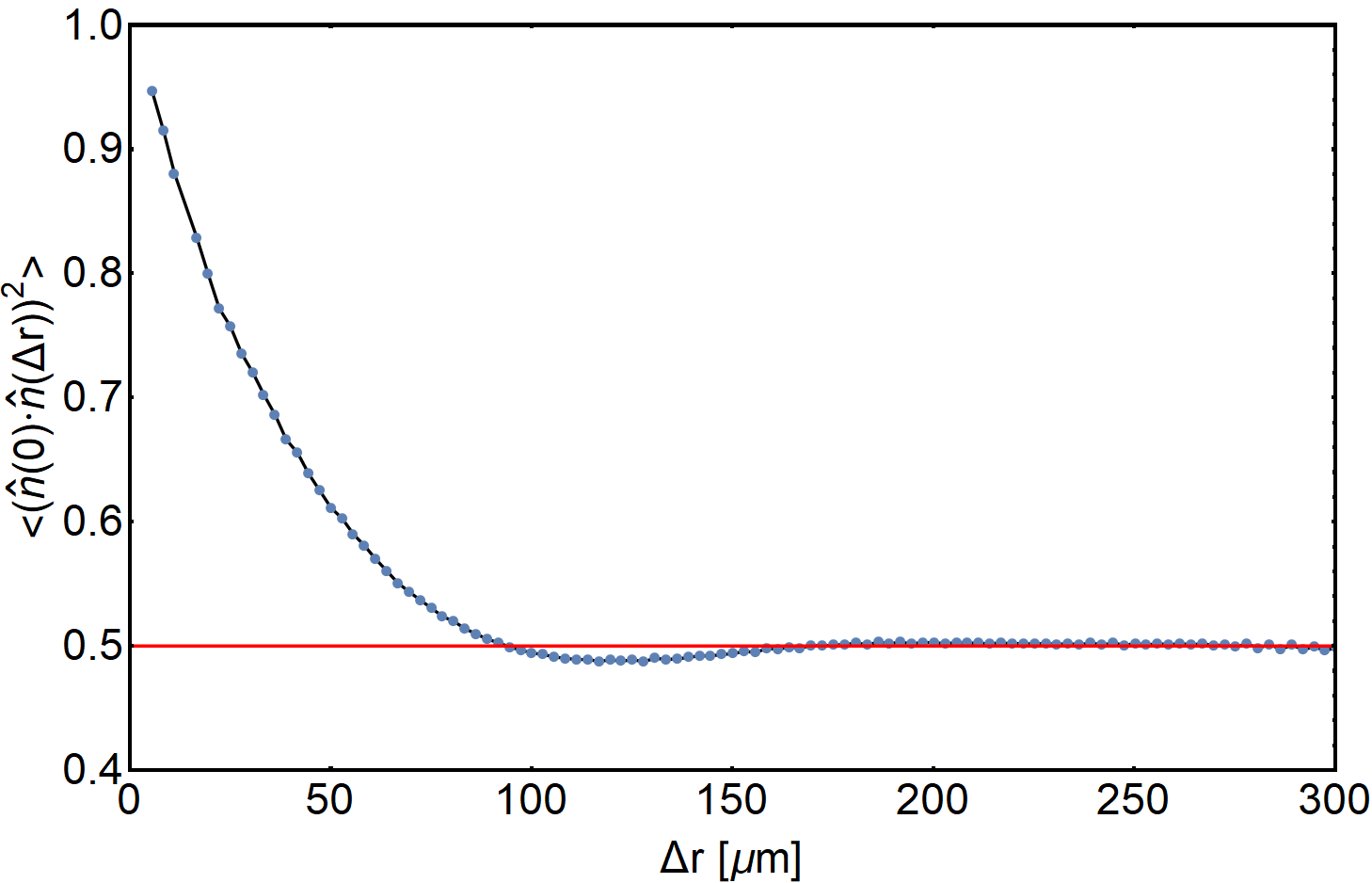}
\caption {The nematic correlation function, defined as the average of the squared dot product of the director at positions separated by $\Delta r$. The results shown are obtained from averaging over a movie of 12 minutes in duration.}
\label{nematiclength}
\end{figure}

Figure~\ref{nematiclength} displays a time and spatial average of the nematic director correlation function, defined as $\langle (\hat{n}(0)\cdot\hat{n}(\Delta r))^2\rangle$, obtained from analyzing a video of an active nematic film spanning 12 minutes and covering a field of view of 570 $\mu$m $\times$ 570 $\mu$m. Due to short-range nematic ordering in the  active nematics, the correlation function is positive at small separations. For random relative orientations of the director, the average value of the correlation function is 0.5, which is the value reached at large separations.   At intermediate separations of approximately 120 $\mu$m, the correlation function drops below 0.5. We identify the the nematic correlation length, shown by the vertical dashed line in the Fig.~5 of the main manuscript, as the location of the minimum of this region.  This distance also roughly equals the average separation between defects ($\sim$130 $\mu$m).  

\section{Simulations}
\subsection{Active Nematohydrodynamics}
The ATP-powered microtubule-bundles/kinesin-complexes mixture at an oil/water interface is simulated as a 2D active nematic film. 
Simulations numerically solve the incompressible active nematohydrodynamics equations of motions~\cite{Marenduzzo2007,thampi2014a} for the local velocity $\vec{u}\left(\vec{r};t\right)$ and nematic order $\tens{Q}\left(\vec{r};t\right)$ fields, which obey the coupled incompressible Navier-Stokes and Beris-Edwards equations
\begin{align}
  \label{eq:dens}
  \del\cdot\vec{u} &= 0  \\
  \label{eq:mom}
  \rho D_t \vec{u} &= \del\cdot\tens{\Pi} - \gamma\vec{u} + \vec{g}\\
  \label{eq:nem}
  D_t \tens{Q} -\tens{\mathcal{S}} &= \Gamma \tens{H}. 
\end{align}
The first of these, $\partial_iu_i=0$, ensures incompressibility, making $\rho$ constant. 

The second equation (Eq.~\ref{eq:mom}) is the Navier-Stokes equation, in which $D_t=\partial_t+u_k\partial_k$ is the material derivative and $\Pi_{ij}$ are the components of the stress. 
The term $g_i$ represents the external force density due to the rotating disk (see Section~\ref{sctn:LB})
The Brinkman term $\gamma u_i$ introduces a friction coefficient $\gamma$ into the film flows~\cite{Thampi2014fric}, which arises from a lubrication approximation for the dissipation within the thin viscous layer between the active film and the bottom boundary~\cite{bruus2008}. 
The friction coefficient screens hydrodynamic flows over distances much larger than $\sim\sqrt{\eta/\gamma\rho}$, where $\eta$ is the viscosity of the film.
We define the strain rate $E_{ij}=\left(\partial_iu_j+\partial_ju_i\right)/2$, vorticity $\mathcal{W}_{ij}=\left(\partial_iu_j-\partial_ju_i\right)/2$ and $\mathcal{Q}_{ij}=Q_{ij}+\delta_{ij}/3$. 
The stresses within film have four contributions~\cite{denniston2004}: 
\begin{enumerate}
  \item The Newtonian viscous stress $\Pi_{ij}^\text{visc}=2\eta E_{ij}$. 
  \item The hydrostatic pressure $\Pi_{ij}^\text{press}=-P \delta_{ij}$, which is taken to be constant. 
  \item The nematic contribution, which itself has symmetric and antisymmetric contributions. 
  \begin{enumerate}
      \item The symmetric terms are $\Pi_{ij}^\text{LC,symm}=2\lambda\mathcal{Q}_{ij} Q_{kl}H_{lk} - \lambda H_{ik}\mathcal{Q}_{kj}-\lambda\mathcal{Q}_{ik}H_{kj} - \partial_i Q_{kl} \frac{\delta F}{\delta\left(\partial_jQ_{lk}\right)}$, where $H_{ij}$ is the molecular field and $F$ is the free energy.
      The alignment parameter $\lambda$ acts to mix upper and lower convective derivative terms. 
      \item The antisymmetric contribution is $\Pi_{ij}^\text{LC,anti}=Q_{ik}H_{kj} - H_{ik}Q_{kj}$.
  \end{enumerate}
  \item The active stress is directly proportional to the liquid crystaline order $\Pi_{ij}^\text{act}=-\zeta Q_{ij}$. The activity $\zeta$ is positive for extensile active nematics, such as the microtubule/kinesin-based films. 
\end{enumerate}

The final equation (Eq.~\ref{eq:nem}) models the evolution of the nematic order tensor. 
The left-hand side of Eq.~ \ref{eq:nem} describes the generalized material derivative accounting for the co-rotational advection $S_{ij}=\left(\lambda E_{ik}+\mathcal{W}_{ik}\right)\mathcal{Q}_{kj} + \mathcal{Q}_{ik}\left(\lambda E_{kj}+\mathcal{W}_{kj}\right) - 2\lambda\mathcal{Q}_{ij}Q_{kl}\partial_ku_l$. 
The $\Gamma H_{ij}$ term describes the relaxation of the order parameter towards the free energy minimum, where $\Gamma$ is a collective rotational diffusivity. 
The molecular field
\begin{align}
    \tens{H} &= -\frac{\delta F}{\delta \tens{Q}} + \frac{\tens{I}}{3}\tr{ \frac{\delta F}{\delta \tens{Q}} }
\end{align}
is symmetric, traceless and is related to the functional derivative of the free energy $F$. 
The free energy is the sum of two contributions $F=\int d^3r\left[f^\text{LdG}+f^\text{el}\right]$, a Landau-de~Gennes bulk free energy density $f^\text{LdG} = A Q_{ij}Q_{ji}/2 + B Q_{ij}Q_{jk}Q_{ki}/3 + C \left(Q_{ij}Q_{ji}\right)^2/4$ and an elastic deformation free energy $f^\text{el} = K\partial_kQ_{ij}\partial_kQ_{ij}$ assuming a single elastic Frank coefficient $K$. 

\subsection{Hybrid Lattice Boltzmann Simulation Details}
\label{sctn:LB}
The equations of motion (Eq.~\ref{eq:dens}-\ref{eq:nem}) are solved using a hybrid lattice Boltzmann/finite difference scheme~\cite{Marenduzzo2007}. 
The Navier-Stokes equation (Eq.~\ref{eq:mom}) is solved using the lattice Boltzmann algorithm on a D3Q15 grid. 
On the other hand, Eq.~\ref{eq:nem} is solved using a finite difference predictor-corrector algorithm. 

Simulations are performed on a two-dimensional domain of size $400\times400$ with periodic boundary conditions for $1.5\times10^{5}$ time steps. 
The discrete space and time steps define the simulations units (su; see Section~\ref{sctn:units}), which are used in this section.
The density is $\rho=1$ and the hydrostatic pressure is $P=1/4$. 
The Landau–de Gennes coefficients used are $A = 0$, $B =-0.3$, and $C = 0.3$, while the rotational diffusivity and alignment parameter are $\Gamma=0.34$ and $\lambda = 0.3$, consistent with our previous works~\cite{Shendruk2018}.
The dynamic viscosity is $\eta = 2/3$. 
The friction coefficient is $\gamma=3\times10^{-4}$.
The Frank elasticity is $K=0.1$, while the activity is $\zeta=5\times10^{-3}$. 
Simulation parameters were chosen to ensure that the disk size in simulation units was large compared to the discrete lattice spacing. 

To account for the hydrodynamic stresses within the nematic film due to the ferromagnetic nickel colloidal disk, we model the rotation resulting from constant external magnetic torque. 
For simplicity , the disk is fixed in place above the film. 
The density of the disk is chosen to be $100$ times that of the film. 
In addition to the magnetic torque, the dissipative drag is $-\zeta_\text{aq}\Omega$, where $\Omega$ is the resulting rotation rate of the disk and $\zeta_\text{aq}=10$ is the rotational drag coefficient of the disk in the aqueous solution. 
The chosen values ensure that the disk's dynamics are over damped. 
The rotation of the disk is hydrodynamically coupled to the active flows within the film. 
We model the effect to be local with coupling only occurring directly below the effective hydrodynamic radius of the disk. 
The hydrodynamic radius $R^\text{H}=25.2$ is taken to be three times the geometric radius of the disk, which accurately captures the experimentally measured position of the maximum average velocity of the film. 
The rotating disk drags the film with a force density proportional to the differences in velocity $\vec{g}=\zeta_\text{d-f}\left[\vec{v}\left(\vec{r},t\right)-\vec{u}\left(\vec{r},t\right)\right]$, where the velocity of each element of the disk is $\vec{v}\left(\vec{r},t\right) = \Omega\left(t\right)\hat{z}\times\vec{r}$, while the force balance produces a net torque on the disk slightly modifying the rotation rate. 

\subsection{Simulation Units}
\label{sctn:units}

\begin{figure}[ht]   
\centering  
%\subfloat{
%\includegraphics[width=0.49\textwidth]{corrLengthA.png}\label{fig:corrL}
%}   
%\subfloat{
%\includegraphics[width=0.49\textwidth]{corrTimeB.png}\label{fig:corrT}
%}  
\includegraphics[width=0.95\textwidth]{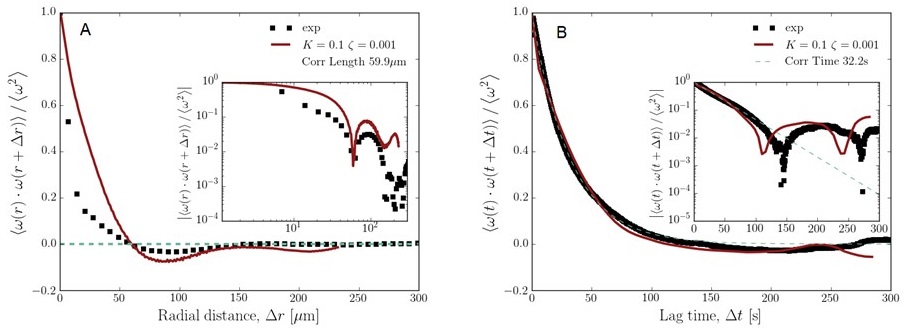}
\caption{The vorticity auto-correlation functions for experiments (black squares) and hybrid-LB simulations (red lines). Panel ({\em A}) shows the same-time spatial auto-correlation function and panel ({\em B}) shows the temporal auto-correlation function. The simulation parameters used are $\gamma=3\times10^{-4}$, $K=0.1$ and $\zeta=5\times10^{-3}$.}
\label{fig:corr}
\end{figure}

In this study care has been taken to match simulation units (su) to the physical conditions of the experimental system. 
Numerical simulations are unitless and appropriate parameters must be chosen to identify possible scalings to convert numerical results to physical units. 
We converted simulation length and time scales to the experiments by comparing decorrelations within the experimental and simulated vorticity fields. 
To determine the length unit conversion we matched the zero crossing point of the spatial vorticity auto-correlation function (Fig.~\ref{fig:corr}; left). 
Experimentally, the correlation length measured in this way is 60 $\mu$m. 
Positive correlation appears more pronounced in the idealized simulations in the near field, as do anti-correlations in the intermediate range. 
In our case, fitting the zero crossing point reproduces the extent of the anti-correlation range. 
This produces the conversion of $1\text{ su}:2.38$ $\mu\text{m}$. 

The unit time conversion is determined by fitting an exponential decay to the short-lag region of the temporal vorticity auto-correlation function (Fig.~\ref{fig:corr}; right). 
The experimental decorrelation is found to be $30$ s. 
This is seen to produce good agreement, both in the short-lag positive correlation and in the intermediate-time anti-correlation. 
This produces the conversion of $43.7\text{ su}:1\text{ s}$. 
These conversions are for a specific experimental activity and oil layer thickness --- we expect them to vary slightly for different experimental realizations.

\section{Image Analysis}
\subsection{Determination of Speed and Vorticity of the Flow in the Active Nematic Films}

To quantify the speed and vorticity of the active nematic films as shown in Fig.~5A of the manuscript, we employed particle image velocimetry (PIV) techniques \cite{PIVLab} to measure the displacement of microtubule bundles in subsections of the film between adjacent video frames. These results were binned and averaged by distance relative to the center of the disk, as determined by the mean location of the disk in the image pair to produce the results shown in Fig.~6A,E of the manuscript.

\subsection{Quantifying Vortex Decay Dynamics} 

\begin{figure}[h]
\centering
\includegraphics[width=0.6\textwidth]{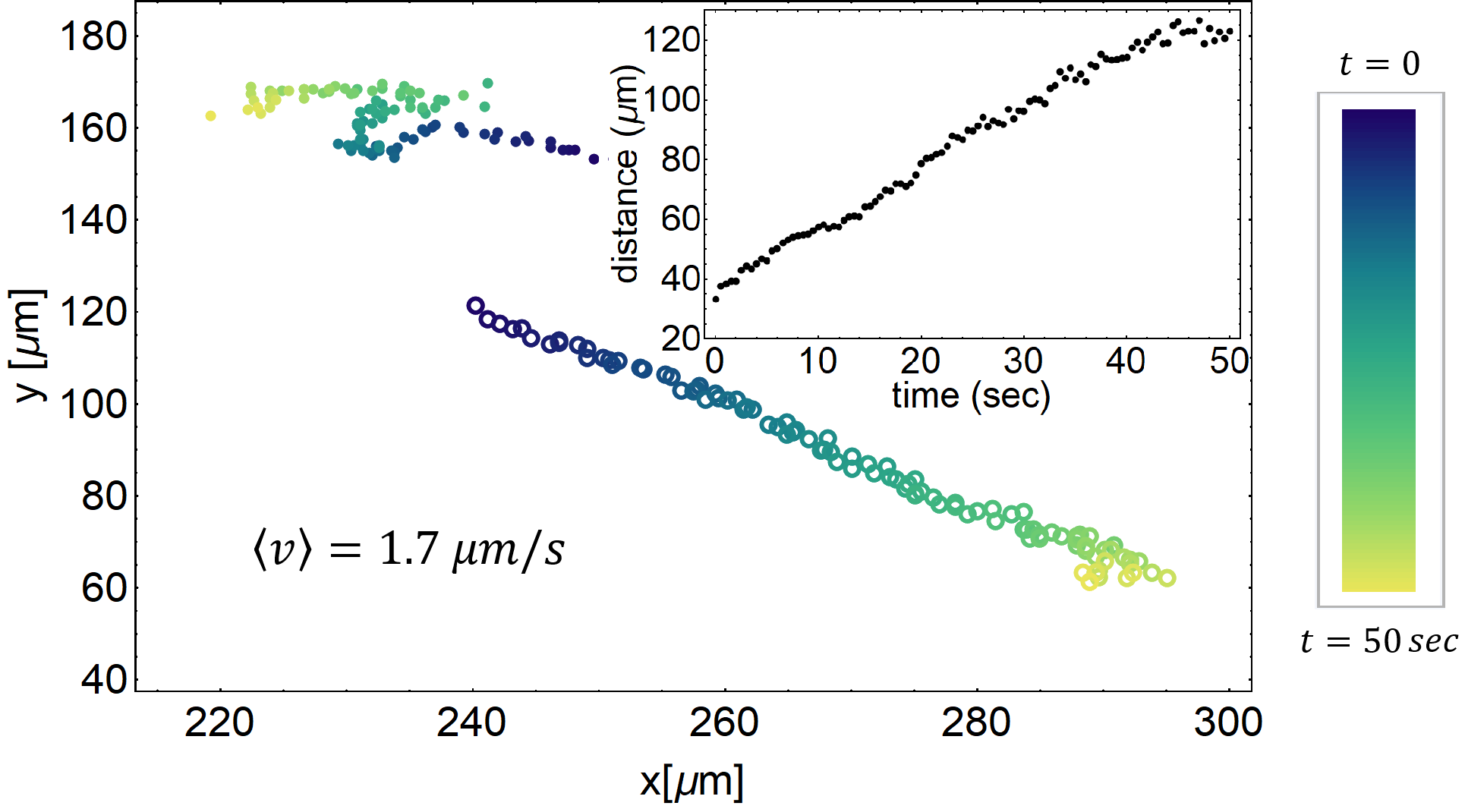}
\caption {Example trajectories of the two vertices of the collapsing elliptical topological structure and the subsequent +1/2 defects. The inset shows their separation versus time. The average speed at which the two features separated was about 1.7 $\mu$m/s. For comparison, the average speed of the +1/2 defects in the film was 1.7 $\pm$ 0.5$\mu$m/s.}
%this is the video 60G_120Hz_30mag_2fps_640_end
\label{VortexCollapse}
\end{figure}

\begin{figure}[h]
\centering
\includegraphics[width=0.6\textwidth]{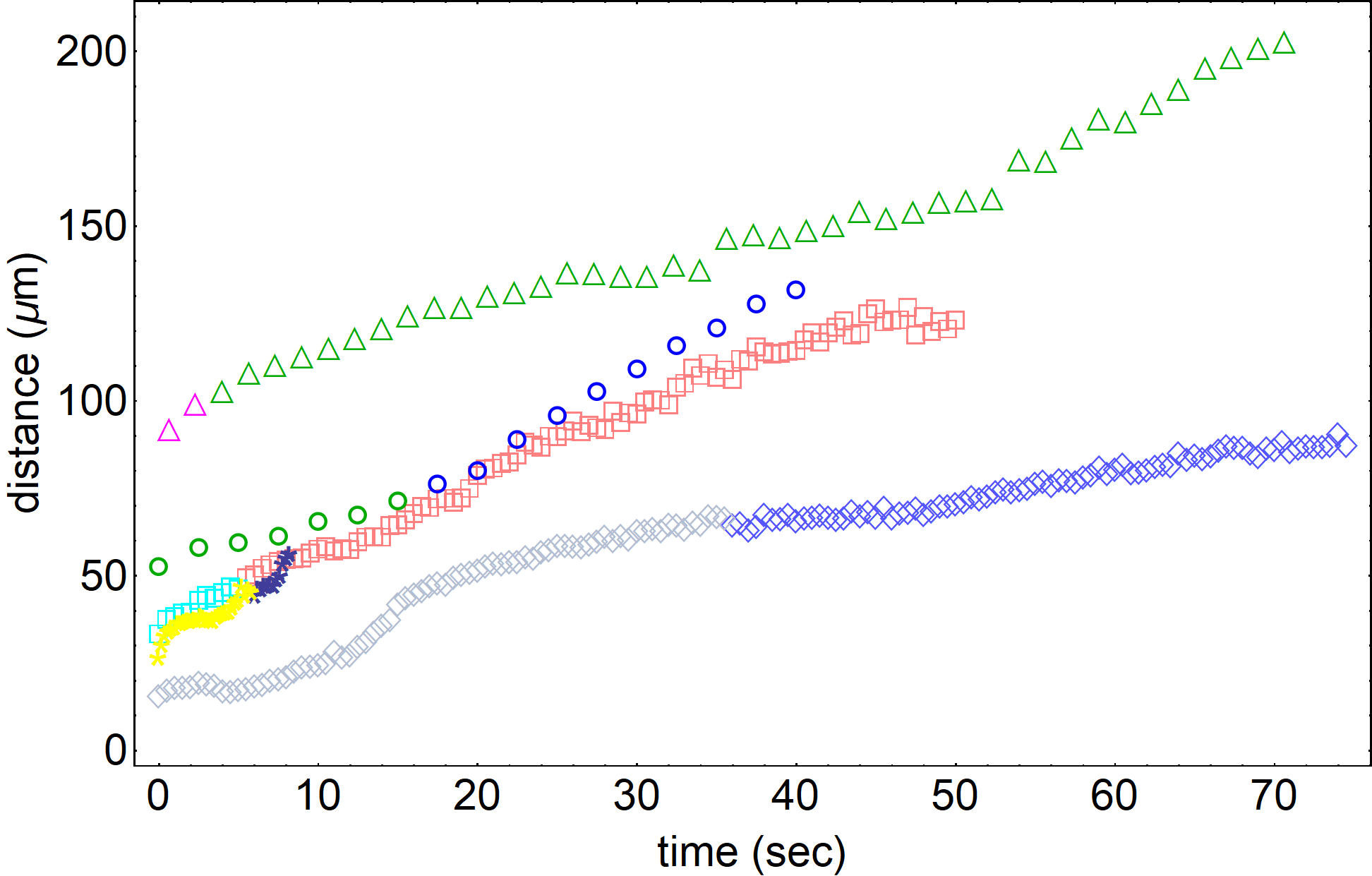}
\caption {Distance between  ends of the collapsing vortices as a function of time for several cases of vortex decay. The change of color of the data points indicates the approximate time at which the elongating +1 topological object became two distinct +1/2 defects.}
\label{VortexCollapseAll}
\end{figure}

 The decay of a topological vortex proceeded by the circularly symmetric director field around the defect core elongating into an elliptical shape in which two +1/2 defects oriented toward each other ({\it i.~e.}, with each defect orientation vector $\psi$ pointing toward the other defect) could quickly be identified.   To characterize the time evolution of the decay process, we tracked this elongation  and then the trajectories of the +1/2 defects by hand using the image processing program ImageJ. An example pair of trajectories is shown in Fig. \ref{VortexCollapse}.  The inset shows the distance between the +1/2 defects as a function of time. (The specific decay time in which the +1 vortex is replaced by two well-defined +1/2 defects is not clearly defined, but  was about 5 s after the start of the elongation into an elliptical director field.)   In Fig.~\ref{VortexCollapseAll} we plot five additional examples  of the separation distance of +1/2 defect pairs produced in topological vortex decays versus time.  The speeds of the +1/2 defects as they propagated away from one another were not measurably different from the typical speeds of the +1/2 defects in the film moving solely as a consequence of the activity.
 
\subsection{Characterization of the Velocity Profile around a Topological Vortex}
 
Individual microtubule bundles in the vicinity of vortices were tracked using custom Python code which, after thresholding, found the centroids of the brightest connected features in each image. These features were then linked using Trackpy's link algorithm~\cite{Trackpy}. The trajectories were hand filtered to select only the region of active nematic with director oriented azimuthally about the vortex. 
The distance of each feature from the vortex core was determined, and the measured speeds of features were averaged over bins of size 10 $\mu$m, resulting in the red data points in Fig.~7. The blue data points in the same figure were produced using a similar method, however, since no topological vortex was present, no filtering of the trajectories was performed. The uncertainties were found by calculating the standard error of the mean of the ensemble average in each bin. Averages of the speeds of each feature were taken rather than an overall average in order to limit the affect of correlations in the data which could skew the mean and underestimate the uncertainty.

\section{COMSOL Calculations of Velocity Profile around a Vortex}

\begin{figure}[h]
\centering
\includegraphics[width=0.5\textwidth]{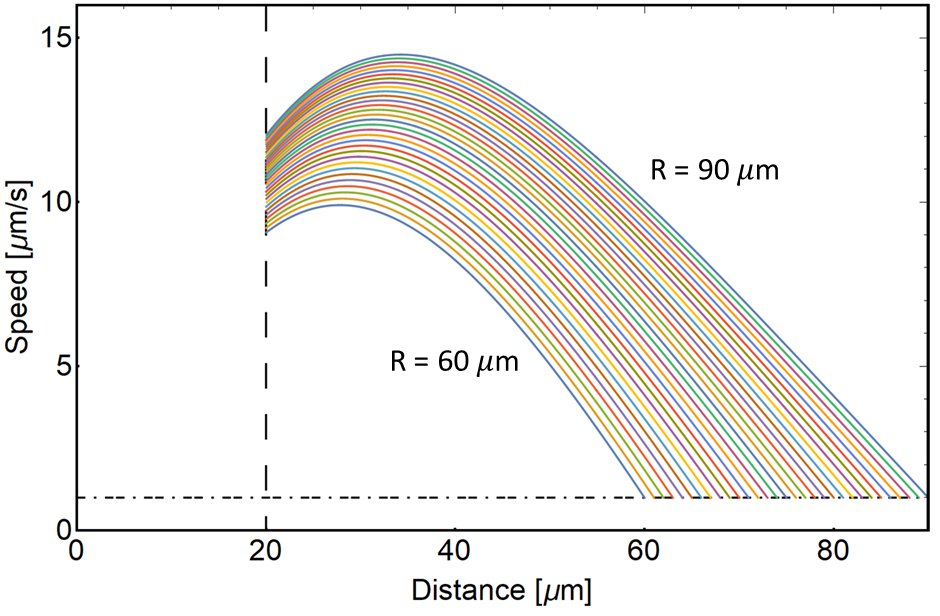}
\caption {Calculated fluid velocity in a circular region of film as a function of distance from the center of the disk for a range of circle radii from 60 to 90$\mu$m in steps of 1$\mu$m. There is a boundary condition of v = 1$\mu$m/s at the outer boundary (shown by the horizontal dot-dashed line). The vertical dashed line represents the radius of the disk. These example velocity profiles were calculated using the same parameters that produced the calculated curve shown in Fig.~7 of the main text.}
\label{SimulationBoundaries}
\end{figure}

As mentioned in the main text, a region of the film in vicinity of the rotating disk experienced a large increase in flow velocity upon formation of a +1 topological vortex (Fig.~7 of the manuscript).  To model this flow profile, we performed hydrodynamic calculations numerically using COMSOL's Computational Fluid Dynamics module. The calculations mimicked the experimental geometry of a disk rotating in a fluid with the viscosity of water (10$^{-3}$ Pa$\cdot$s) at a set height (35 $\mu$m) above a film of thickness 300 nm and viscosity $\eta_f$, which was varied in the calculations to optimize the agreement with the experimental velocity profile.  The space below the film was similarly treated as a fluid with viscosity matching that of the oil in the experiments (10$^{-3}$ Pa$\cdot$s).  Several thicknesses for the oil layer were tested in the calculations, and the resulting velocity profile in the film was insensitive to the layer thickness as long as it was greater than $\approx 1$ $\mu$m.  Since the region of enhanced film velocity in the experiments typically coincided with the region of active nematic centered on the disk in which the director field was oriented in the azimuthal direction, we modeled the film as a circle of radius $R$ matching the size of this region.  As a boundary condition we set the velocity at the edge of the circle equal to the experimentally measured speed of the surrounding active nematic (approximately 1 $\mu$m/s).  Over the lifetime of a vortex, the region of azimuthally oriented director field typically fluctuated.  For example, for the vortex analyzed to produce the velocity profile in Fig.~7 $R$ varied between approximately $R = $ 60 $\mu$m and $R = $ 90 $\mu$m during the measurements.  Therefore, we calculated velocity profiles in films with $R$ ranging from 60 to 90 $\mu$m and took the average of these curves to compare with the data.  This process was repeated for different values of $\eta_f$ to find the optimal agreement with the experimental data.  For instance, Fig.~\ref{SimulationBoundaries} shows a set of curves that were averaged to produce the calculated curve shown in Fig.~7.

%%% Add this line AFTER all your figures and tables
\FloatBarrier

\movie{Microscopy video of the merger of two +1/2 defects into a +1 topological vortex in an active nematic film due to stresses produced by a rotating disk. The images in Fig.~2 are taken from this video.}

\movie{Lattice-Boltzmann simulation of the merger of two +1/2 defects into a +1 topological object. The images in Fig.~3 are from this simulation.}

\movie{Microscopy video of the decay of a topological vortex into two +1/2 defects. The images in Fig.~4A-D are taken from this video. }

\movie{Lattice-Boltzmann simulation of the decay of a +1 topological object into two +1/2 defects. The images in Fig.~4E-H are from this simulation.}

\movie{Microscopy video of active nematic films in the presence of a rotating disk during a period in which no topological vortex forms. Note that the +1/2 defects in proximity to the disk tend strongly to circulate anti-clockwise around the disk.}

\bibliography{topological_transitions_citations_8_29_19.bib}